\documentclass[a4paper,11pt]{article}
\usepackage[margin=1.3cm,a4paper]{geometry}
\usepackage{amssymb}
\usepackage{amsmath}
\usepackage{bm}
\usepackage{graphicx}
\graphicspath{ {images/} }
\usepackage{subcaption}

\usepackage{setspace}
\usepackage[utf8]{inputenc}
\usepackage{url}
\usepackage{color}

\usepackage[]{apacite} 

\newcommand*\mean[1]{\bar{#1}}

\begin{document}


\pagenumbering{gobble}

\begin{center} 

    \centering
    \vspace{1.5cm}

    {\uppercase{\Large Modelling Antimicrobial Prescriptions in Scotland: \\ a spatio-temporal clustering approach \par}}

\end{center}

\vspace{0.5cm}

\pagenumbering{arabic}

Antonia Gieschen\textsuperscript{a}* \\
\textit{Antonia.Gieschen@ed.ac.uk}\\

Jake Ansell\textsuperscript{a} \\
\textit{J.Ansell@ed.ac.uk}\\

Raffaella Calabrese\textsuperscript{a} \\
\textit{Raffaella.Calabrese@ed.ac.uk}\\

Belen Martin-Barragan\textsuperscript{a}\\
\textit{Belen.Martin@ed.ac.uk}

\vspace{0.1cm}

{\footnotesize\textsuperscript{a} University of Edinburgh Business School, 29 Buccleuch Place, Edinburgh, EH8 9JS , United Kingdom}

\section*{Abstract}

In 2016 the British government acknowledged the importance of reducing antimicrobial prescriptions in order to avoid the long-term harmful effects of over-prescription. Prescription needs are highly dependent on factors that have a spatio-temporal component, such as the presence of a bacterial outbreak and the population density. In this context, density-based clustering algorithms are flexible tools to analyse data by searching for group structures.  The case of Scotland presents an additional challenge due to the diversity of population densities under the area of study. We present here a spatio-temporal clustering approach for highlighting the behaviour of general practitioners (GPs) in Scotland.\\
Particularly, we consider the density-based spatial clustering of applications with noise algorithm (DBSCAN) due to its ability to include both spatial and temporal data, as well as its flexibility to be extended with further variables. We extend this approach into two directions. For the temporal analysis, we use dynamic time warping to measure the dissimilarity between warped and shifted time series. For the spatial component, we introduce a new way of weighting spatial distances with continuous weights derived from a KDE-based process. This makes our approach suitable for cases involving spatial clusters with differing densities, which is a well-known issue for the original DBSCAN. We show an improved performance compared to both the latter and the popular k-means algorithm on simulated, as well as empirical data, presenting evidence for the ability to cluster more elements correctly and deliver actionable insights.

\section*{Keywords}

OR in health services -- Multivariate statistics -- Cluster analysis -- Spatio–temporal data

\newpage

\section{Introduction}
\label{Introduction}


The need to identify unwarranted variation and extreme behaviour in drug prescribing was highlighted recently in 2018 by Catherine Calderwood, the Chief Medical Officer for Scotland, in her report titled `Practising Realistic Medicine': ``Unwarranted variation [...] is variation in healthcare that cannot be explained by need, or by explicit patient or population preferences. Recognising unwarranted variation is of vital importance because it allows the identification of: Underuse of higher value interventions – i.e. under treatment. Over use of interventions which should be used less frequently. Over use of interventions which may result in harm.'' \cite{healthcare2018}). Motivated by this, our empirical application leads to a better understanding of prescription behaviour in Scotland regarding its temporal trends, how it differs across regions and differences between GP practices. The areas showing different behaviour patterns can present evidence for distinct policy or communication strategies. Regions with higher demand for specific medications could have a higher rate of underlying health issues. Highlighting extreme behaviour in the form of outliers as well as extreme behaviour within clusters can also offer valuable insights into the behaviour of specific GPs or specific, relatively small-scale areas. In a broader sense, our findings can be used in connection with a disease recognition and, ideally, prevention system. The goal of our research is to inform more effective prescribing by identifying unwarranted variation, to minimise harm, and reduce waste for the National Health Service (NHS) in Scotland. The specific focus on Co-Amoxiclav or Amoxicillin is partially driven by the consideration of the UK 5 Year Antimicrobial Resistance Strategy 2013-2018, which had the aim of: ``optimising prescribing practice through implementation of antimicrobial stewardship programmes that promote rational prescribing [...]'' \cite{department2016}.\\


Using data obtained from the NHS Scotland Open Data platform, we can explore the issue of antimicrobial prescriptions. The goal is to identify groups or outliers that reflect positive or negative prescribing behaviours. As the groups are unknown, this suggests the use of clustering. We expect antimicrobial prescriptions to follow a yearly cycle \shortcite{durkin2018}, calling for the need of considering the time element. Spatial aspects reflect the role of health and socio-demographic effects \shortcite{kjaerulff2016, molter2018}. Our choice is therefore to explore the application of spatio-temporal clustering in this context. Specifically, we analyse the characteristics of our proposal on antimicrobial prescriptions in Scotland, by using a density-based spatio-temporal clustering algorithm to group together (cluster) general practitioners (GPs) with respect to their prescription volume behaviour over time and over space in Scotland. The use of spatial and spatio-temporal clustering techniques to analyse diseases or drug prescriptions has been subject to extensive discussion in the literature \shortcite{anderson2016, blangiardo2016}. \citeA{katz2010} analyse prescriptions on an individual patient level whereas we use data aggregated at GP level. Antimicrobial prescriptions in particular have also been subject of research \cite{petersen2007}. When analysing the spatial distribution of diseases, other factors are also considered and can be included in our approach, for example socio-demographic data \shortcite{kjaerulff2016}. \citeA{molter2018} analyse antibiotic prescription behaviour in England using a hotspot analysis, and connect their findings to the socio-demographic situation in these areas. Differences between urban and rural areas in the context of health have also been subject of research. \shortciteA{cairns2011} use simulation to analyse first-responder schemes for cardiac arrests in Northern Ireland, discussing spatial complexities that come with it for example driven by differences between urban and rural areas.\\


Our reason to choose a clustering algorithm for this analysis lies in their many advantages when analysing complex data sets to search for groups, for example by eliminating the need for pre-labelled data and the pre-knowledge about classes. Clustering algorithms can be classified depending on the way they assign the data points to their groups, for example by partitioning, building a hierarchy in the form of a dendrogram, or looking for density \shortcite{han2011}. They all group together data points into clusters based on their similarity and the choice of this similarity measure differs depending on the data type. Many clustering methods, especially for spatial clustering, use distance measures to calculate the dissimilarity between objects. Some examples belonging to this group include the k-means algorithm \shortcite[pp.451-454]{han2011}, the ``Clustering Large Applications based on RANdomized Search" (CLARANS) algorithm \cite{ng2002} and the density-based algorithm DBSCAN \cite{ester1996}. Density-based clustering algorithms, such as DBSCAN, use the distance measures to assess the density of points in a region to determine whether a cluster exists in that location. These algorithms, however, suffer from a drawback. Varying densities in spatial dimensions can lead to the algorithm not being able to identify clusters in both very dense and not dense areas. In a practical context, the ability to handle varying densities becomes relevant when comparing urban and rural regions due to different spatial prevalence across areas. This is the case, for example, in Scotland, where highly dense urban areas are concentrated in the South while the centre, the North and the islands remain quite unpopulated. In order to solve this, related research usually focusses on using a nearest neighbour approach to determine density, such as DECODE \shortcite{pei2009} or the use of shared nearest neighbours \shortcite{ertoez2003}. \citeA{ertoez2003} propose to define the similarity between two points based on the number of shared nearest neighbours. For this solution, however, one needs to decide the number of nearest neighbours which should be considered by the algorithm. This choice can strongly impact the clustering results. \\

Clustering of time series often uses transformations of the time series by representing them in a stochastic way for example with Markov Models (model-based approaches) or, alternatively, features (feature-based approaches). Then subsequently the clustering is conducted with the extracted model parameters or features \cite{deAngelis2014}. Alternatively, for shape-based approaches, clustering algorithms compare the raw, untransformed, time series by trying to match them by stretching or shifting them \shortcite{aghabozorgi2015}. In all three cases, one needs to determine the measure of similarity between time series. This can be difficult, because effects may not be synchronised, but delayed or shifted. One example is the increase in the number of cases of a disease like the flu, which might occur in different regions at separate times. The challenge of shifted time series has been discussed in the literature and lead to the introduction of methods such as dynamic time warping (DTW) \cite{serra2014}. DTW calculates the dissimilarity of two time series by minimising the 'cost' needed to match them. While it is a popular method in time series clustering, to the best of our knowledge it has not been used in connection with spatial data as input for the DBSCAN algorithm. \\


Our proposed method is able to handle both of these identified challenges, namely clusters of varying densities and shifted time series. In contrast to the nearest neighbour solutions in the literature \shortcite{kriegel2011,ertoez2003}, our approach is continuous, covering points with a smooth function, without considering a specified number of neighbours. The main contribution of this work can be found in its methodological novelty of using continuous density estimates around spatial data points to weight their distances to one another by deriving weights through a logistic function. This idea of using the density measures derived via kernel density estimation (KDE) \cite{silverman1986}, which we introduce here in a DBSCAN algorithm, is very flexible. It can be implemented for a variety of density-based clustering algorithms which consider spatial information in the form of a distance matrix. Examples for this include, but are not limited to, DBSCAN \shortcite{ester1996} and many of its extensions. In this paper, we implement our approach using ST-DBSCAN, a spatio-temporal version of the density-based clustering algorithm DBSCAN \cite{birant2007}. By weighting the spatial distance matrix with continuous distance weights derived from a KDE-based process, we are able to make our approach suitable for spatial clusters with varying densities. The algorithm makes use of KDE in order to formulate an area's density \shortcite{terrell1992, silverman1986}. While previous work formulates a region's density based on a point's nearest neighbours, this approach provides us with a smooth density estimate in every location of our spatial map. The specific advantage of our approach lies in its increased suitability for real-life applications in which spatial clusters exhibit varying densities and the global neighbourhood radius used by DBSCAN to identify clusters is not accurate, as can be seen in our application. Furthermore, the methodological contribution of our work is usage of KDE for scaling which contributes to the area of local scaling and dealing with varying densities for clustering. This is of special interest for organisations and companies which are also looking for ways of graphically presenting their results, making them more accessible.
Furthermore, regarding the temporal aspect, this approach is combined with a flexible way of measuring temporal dissimilarity by employing DTW. This results in a spatio-temporal clustering algorithm suitable for the analysis of time series which are spatially unequally distributed.\\ 

We test our approach against k-means as the most commonly used clustering algorithm, as well as against the original DBSCAN algorithm. Our results show an improved performance regarding the number of correctly clustered elements when compared with both these algorithms. When tested on 3000 simulated spatial data points, our proposed method reduces the number of incorrectly clustered elements from 1211/3000 (k-means) and 308/3000 (DBSCAN) to 225/3000. At the same time, the number of outliers is reduced from 256/3000 (DBSCAN) to 183/3000. Applied on spatio-temporal empirical data, the proposed method is able to cluster both a dense and a sparse area simultaneously, while the original ST-DBSCAN algorithm either over-smoothed them or identified a very high number of small clusters. \\

We provide background information on both clustering and KDE in Section \ref{background}, while Section \ref{methodology} describes the proposed methodology in detail. Our approach is tested against the original DBSCAN to show its improvements, as well as against k-means as a common baseline algorithm, in a simulation in Section \ref{simulation}. In Section \ref{application} our method is applied in a case study of GP prescription behaviour in Scotland to demonstrate its performance for spatio-temporal clusters with varying densities and shifted time series. The paper concludes and offers suggestions for further research in Section \ref{conclusion}.


\section{Literature review}
\label{background}

Clustering is an unsupervised machine learning technique for grouping data points. In contrast to supervised classification, it finds and groups data structures according to their (dis)similarity instead of requiring a training set of pre-labelled examples \cite{han2011, kaufman2009}. Clustering applications are broad and range from biology \shortcite{kiselev2019} to economics \cite{dupor2017}, and pattern recognition \shortcite{schroff2015} to medicine \cite{dougherty2002}. The choice of algorithm depends on the type and amount of data, as well as on the objective. Well-known algorithms include partitioning methods such as k-means, which optimises an objective function calculating the intra-cluster and inter-cluster distances. There are numerous adaptations which, for example, respect upper and lower bounds for the number of clusters \shortcite{borgwardt2017}. Instead of finding one outcome, hierarchical approaches produce tree-like structures called dendrograms \cite{han2011}. Probabilistic approaches, including mixture models, assume that each cluster is one of a mixture of subpopulations \shortcite{mai2018, liao2005}. Graph clustering looks for clusters of vertices in graphs in such a way that there are many edges within a cluster and relatively few between them \shortcite{benati2017, schaeffer2007}. One group of algorithms in this area are spectral methods, where an eigenvector or a combination of eigenvectors is used for computing the similarity of clusters \cite{nascimento2011, schaeffer2007}, while another approach is to formulate a min-sum problem which minimises the sum of distances between points in a cluster \cite{hassin2010}.\\

Spatial clustering describes using cluster analysis to cluster data points in geographic regions. For example, \shortciteA{coll2014} use hierarchical clustering to investigate patterns of vehicle collisions within an identified hot spot area in Northern Ireland. In the literature spatial clustering is sometimes treated as a special case due to the assumptions made in the wider area of, for example, spatial statistics, namely interdependency structure between locations \cite{menafoglio2017, cressie1992}. This means that we assume that locations which are close influence each other more strongly than those further away from each other. There have been efforts to capture this dependency structure through copulas based clustering, which can be interpreted as a sub-category of model-based clustering in which copula information is used to derive the cluster composition \shortcite{diLascio2017}. \shortciteA{disegna2017} use copulas to capture spatial dependency structures in their dissimilarity matrix, which they then feed into a fuzzy clustering algorithm. The latter does not allocate each object to one and only one cluster, but instead allows for membership in multiple clusters by assigning membership degrees \cite[pp.242-245]{everitt2011}. \shortciteA{marbac2017} use a mixture of Gaussian copulas for clustering mixed data, commenting on the advantage of being able to interpret the dependency structure through the copula.\\ 

Approaches commonly used for spatial data are density-based methods such as the DBSCAN algorithm \cite{ester1996}, which is explained in more detail in Section \ref{methodology}. This algorithm looks for high-density neighbourhoods of points which are similar or close to each other, and segregated by areas of lower density. Another method using the concept of neighbours is CLARANS \cite{ng2002}, which is able to cluster not only points but also polygonal objects. There are two main advantages of density-based methods. The first is their ability to find clusters of arbitrary shape. The second is the fact that many density-based methods do not require the number of clusters as an input as does, for example, k-means.\\

There are many adaptations to DBSCAN. These include the constraint-based C-DBSCAN \shortcite{ruiz2007}, which uses pre-knowledge about group memberships, sampling-based IDBSCAN \cite{borah2004}, which is especially suitable for very large spatial databases, and the spatio-temporal ST-DBSCAN \cite{birant2007}. Another extension on the original DBSCAN algorithm is ``Ordering Points to Identify the Clustering Structure'' (OPTICS) \shortcite{ankerst1999}, which produces not a clustering but an ordering of the database that represents the underlying clustering structure. It is ordered in such a way that spatially close points are neighbours in the ordering, and objects in dense clusters are processed first. Besides being a good tool for visualisation purposes, OPTICS's main advantage is its ability to handle varying densities by analysing the neighbours for each point. Similarly, ``Locally Scaled Density Based Clustering'' (LSDBC) \cite{biccici2007} also uses neighbouring points, in this case to determine the local density maxima in regions which act as cluster centres. It aims to identify clusters within background noise and is thought to be more robust towards parameter changes. LSDBC connects density regions until the overall density falls below a pre-defined threshold. \citeA{rodriguez2014} propose ``Clustering by Fast Search and Find of Density Peaks'' (CFSFDP), which identifies cluster centres by defining them as points with a high local density and a high distance to other such centres. Subsequent work proposes approaches to identify the optimal values for these parameters \shortcite{mehmood2016} and fuzzy implementations of the algorithm \shortcite{bie2016}. \\

Density-based clustering methods are applied to a broad range of research areas. \citeA{gomide2011} use Twitter data to analyse a Dengue fever epidemic. The authors use ST-DBSCAN to cluster cities that show a similar incidence rate of Dengue fever occurring during the same period of time, exploring whether Twitter can be used to predict the number of Dengue fever cases. For this purpose they compare the number of cases reported by official statistics and the number of tweets about the disease during the same time period. This enables them to develop a Dengue fever surveillance tool based on four dimensions, namely volume, location, time, and public perception, showcasing the real life applicability of their approach. In addition, \shortciteA{anbaroglu2014} comment on the use of spatio-temporal clustering to detect non-recurrent traffic congestion in London, as clusters of slowly-moving vehicles can cause congestion in an urban road network.\\

DBSCAN and adaptations like ST-DBSCAN face problems when trying to detect clusters with varying point densities due to the use of global parameters when looking for close neighbouring points that indicate existing clusters. An example for a situation like this can be seen in Scotland, where the population density varies across the region. This issue has been addressed in the literature, for example by \shortciteA{kriegel2011}. They explain how density-based clustering methods often label points in less dense areas as noise or outliers when there are regions of high density separated by areas of lower density. As potential solutions to this problem, they propose hierarchical approaches of ordering points such as OPTICS \cite{ankerst1999}, as well as nearest neighbour approaches \cite{ertoez2003, pei2009}.\\

\citeA{birant2007} discuss varying densities in their paper on ST-DBSCAN and outline the problem of identifying actual noise in a scenario of high-density areas surrounded by regions of lower density. They propose assigning each cluster a density factor, calculated by looking at the maximum and minimum distances within a respective cluster. This single factor might, however, not reflect the various different density regions which are present across each cluster, especially if the area described by one factor is relatively large.\\

Similarly, \citeA{zelnik2005} discuss the idea of local scaling in the context of their self-tuning spectral clustering algorithm. Based on this, \citeA{biccici2007} combine this local scaling with an adapted density-based clustering algorithm. Their density estimation, however, is based on a $k$-nearest neighbour approach. There are two aspects to this approach which can be considered drawbacks. First, one has to decide the number of considered neighbours $k$, which is not the case for our approach. Secondly, in contrast to our method, the approach by \citeA{biccici2007} does not produce a continuous density estimation map.\\

Clustering can also be employed in time series analysis \cite{deAngelis2014, serra2014}. For clustering such data, emphasis in the literature has been put on the selection of the most suitable dissimilarity measure. In general, there are four ways of measuring the dissimilarity between time series, as reviewed by \citeA{serra2014}. Lock-step methods such as the Euclidean distance measure similarity by comparing values at the same time steps in each series, their main advantage being computational simplicity \shortcite{xing2012}. Feature-based methods employ the same approach, but work on a transformation of the time series. This can, under some circumstances, improve the accuracy of the dissimilarity compared to the raw Euclidean distance and is, therefore, commonly used \shortcite{serra2014, montani2006}. \citeA{inniss2006} introduces an approach for seasonal clustering of time series applied to weather and aviation data. They form clusters of contiguous months with similar weather conditions, using the mean value of a month, as measured over several years, or the difference between distributions for each month, over a number of years. Model-based approaches assume an underlying model generating the time series and measure dissimilarity based on the model parameters. For example, mixture  models are used by \shortciteA{povinelli2004}, who comment that the method is able to perform well on a range of application areas with minimal input tuning. \shortciteA{dias2015} present an approach which is able to account for time-constant unobserved heterogeneity and hidden regimes within time series by using hidden Markov models. Elastic methods compare the time series by their overall shape over time instead of their absolute values measured at each time step. Dynamic Time Warping (DTW) aims to optimally align the two time series by minimising a warping cost. It has been used in various contexts, including for incomplete medical time series \shortcite{tormene2009} and to measure the dissimilarity between time series in combination with fuzzy clustering \shortcite{izakian2015}. ``Edit Distance on Real Sequence'' (EDR) calculates dissimilarity as the number of edit operations necessary for transforming one time series into the other \shortcite{chen2005}. $k$-Shape and $k$-MS are iterative time series clustering algorithms which use shape-based distance (SBD) as their dissimilarity measure \cite{paparrizos2017}. The authors comment on the fact that their algorithm is able to outperform DTW, especially regarding computational time and the fact that their approach requires no parameter tuning.\\


This paper makes the following two contributions. The first is related to DBSCAN's shortcomings when confronted with varying point densities. Existing literature has tackled this problem with nearest neighbour approaches \cite{kriegel2011, pei2009, biccici2007, zelnik2005, ertoez2003} or by assigning density factors to clusters \cite{birant2007}. To the best of our knowledge, no continuous approach using KDE has been used for this purpose so far. The second gap is the usage of a time series dissimilarity measure which is able to handle shifted or warped times series. Here, we tackle this by employing DTW with the spatio-temporal clustering algorithm ST-DBSCAN.


\section{Methodology}
\label{methodology}
 
The objective of this paper is to enable the spatio-temporal clustering of a large amount of data points forming clusters with varying densities, each of which is associated with a univariate time series. Our approach can be used in connection with a variety of clustering algorithms but will be demonstrated by ST-DBSCAN allowing spatio-temporal clustering \cite{birant2007}. This algorithm is chosen due to its computational efficiency and its flexibility with regard to the addition of more variables alongside the temporal and spatial information. \\
In order to tackle ST-DBSCAN's inability to handle unevenly distributed spatial points, our approach manipulates the input spatial distance matrix using weights based on the location density derived through KDE. This makes the spatial distances of locations with varying point densities comparable to each other. In doing so, the use of a global spatial radius parameter, which identifies other cluster members belonging to the same group, becomes possible.

\subsection{Clustering using DBSCAN and ST-DBSCAN}
\label{method:dbscan}

DBSCAN operates by visiting a random data point and looking for neighbouring points within a pre-defined maximum spatial radius $\varepsilon_{1}$. \citeA{ester1996} define the $\varepsilon$-neighbourhood of a point $p$ as $N_{\varepsilon}(p)$, with 
\begin{equation}\label{eq:1}
N_{\varepsilon}(p) = \{q \in D \ | \ \mathrm{dist}(p,q) \leq \varepsilon\}.
\end{equation} 

If $N_{\varepsilon}(p) \geq \eta$, that means the number of neighbouring points in the neighbourhood described by Equation (\ref{eq:1}) is more than a pre-defined number $\eta$, the respective area is considered a dense area, and a cluster is formed. In that case $p$ is called a \textit{core point}. Points which are subsequently assigned as members of this cluster without fulfilling the condition are called \textit{border points}. Points outside of clusters are labelled as \textit{noise}. \\

In order to form the clusters, \citeA{ester1996} define a number of concepts regarding the reachability of points. To capture both border and core points, these concepts are \textit{directly density-reachable}, \textit{density-reachable} and \textit{density-connected}.\\

A point $q$ is \textit{directly density-reachable} (DDR) from a point $p$ if two conditions hold true, namely

\begin{figure}[h]
\center
\includegraphics[width=0.4\textwidth]{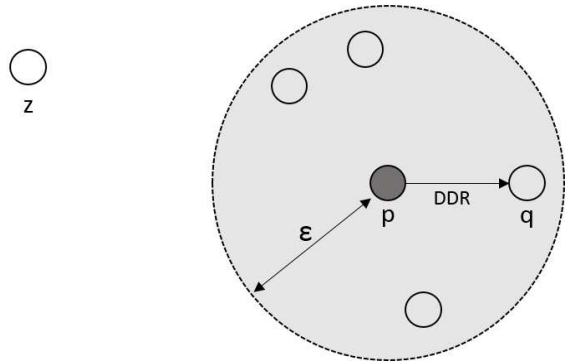}
\caption{The concepts of core and border points and direct density-reachability with $\eta = 4$. Point $p$ is a core point (dark grey), as $|N_{\varepsilon}(p)| \geq 4$, shown as a shaded grey area with radius $\varepsilon$. Point $q$ is directly density-reachable (DDR) from point $p$ as $q \in N_{\varepsilon}(p)$ and $p$ is a core point. In this case, $q$ is a border point (light grey) as $|N_{\varepsilon}(q)| \leq \eta$ but $q$ belongs to the cluster around $p$ which is beginning to form here. $z$ is a noise point as it is outside of all core point neighbourhoods.}
\label{fig:dbscan1}
\end{figure}

\begin{enumerate}
\item $q \in N_{\varepsilon}(p)$ and
\item $|N_{\varepsilon}(p)| \geq \eta$,
\end{enumerate}

with the second condition being called the \textit{core point condition}. This means that $q$ must be in the $\varepsilon$-neighbourhood of $p$, and $p$ must be a core point. This is visualised in Figure \ref{fig:dbscan1}.

\begin{figure}[h]
\centering
\begin{subfigure}{.3\textwidth}
  \centering
  \includegraphics[width=.9\linewidth]{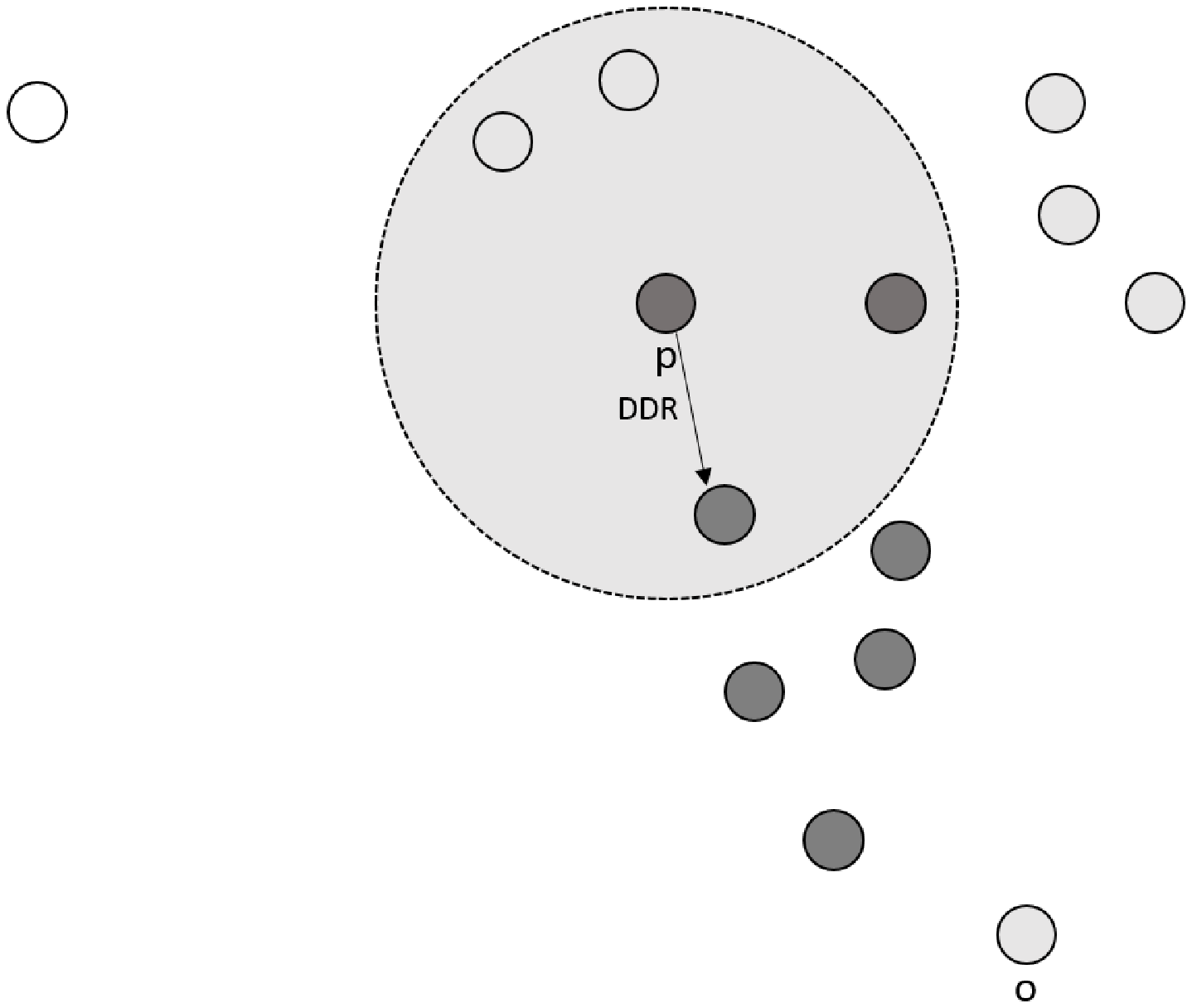}
  \label{fig:sub1}
\end{subfigure}%
\begin{subfigure}{.3\textwidth}
  \centering
  \includegraphics[width=.9\linewidth]{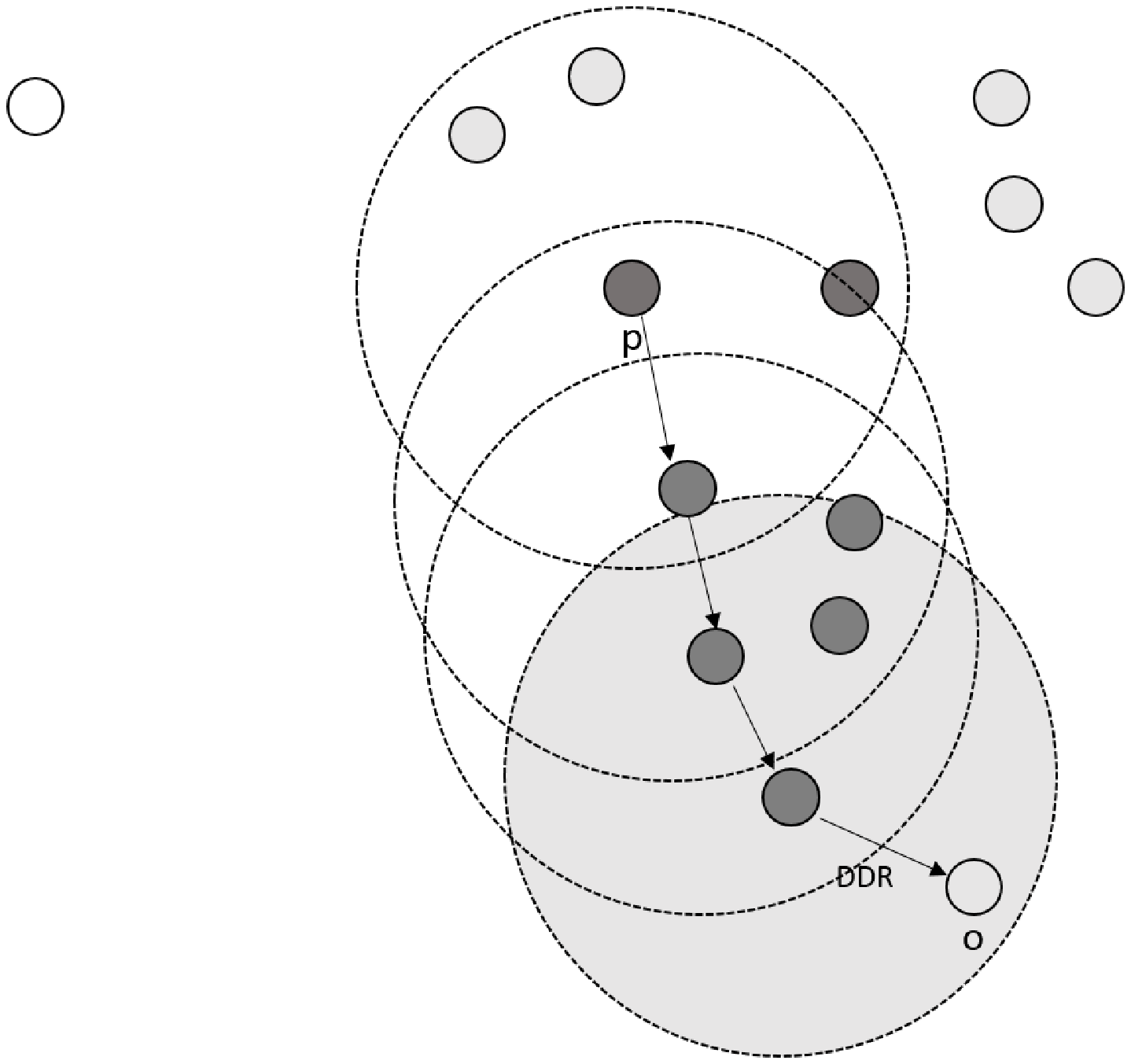}
  \label{fig:sub2}
\end{subfigure}%
\begin{subfigure}{.3\textwidth}
  \centering
  \includegraphics[width=.9\linewidth]{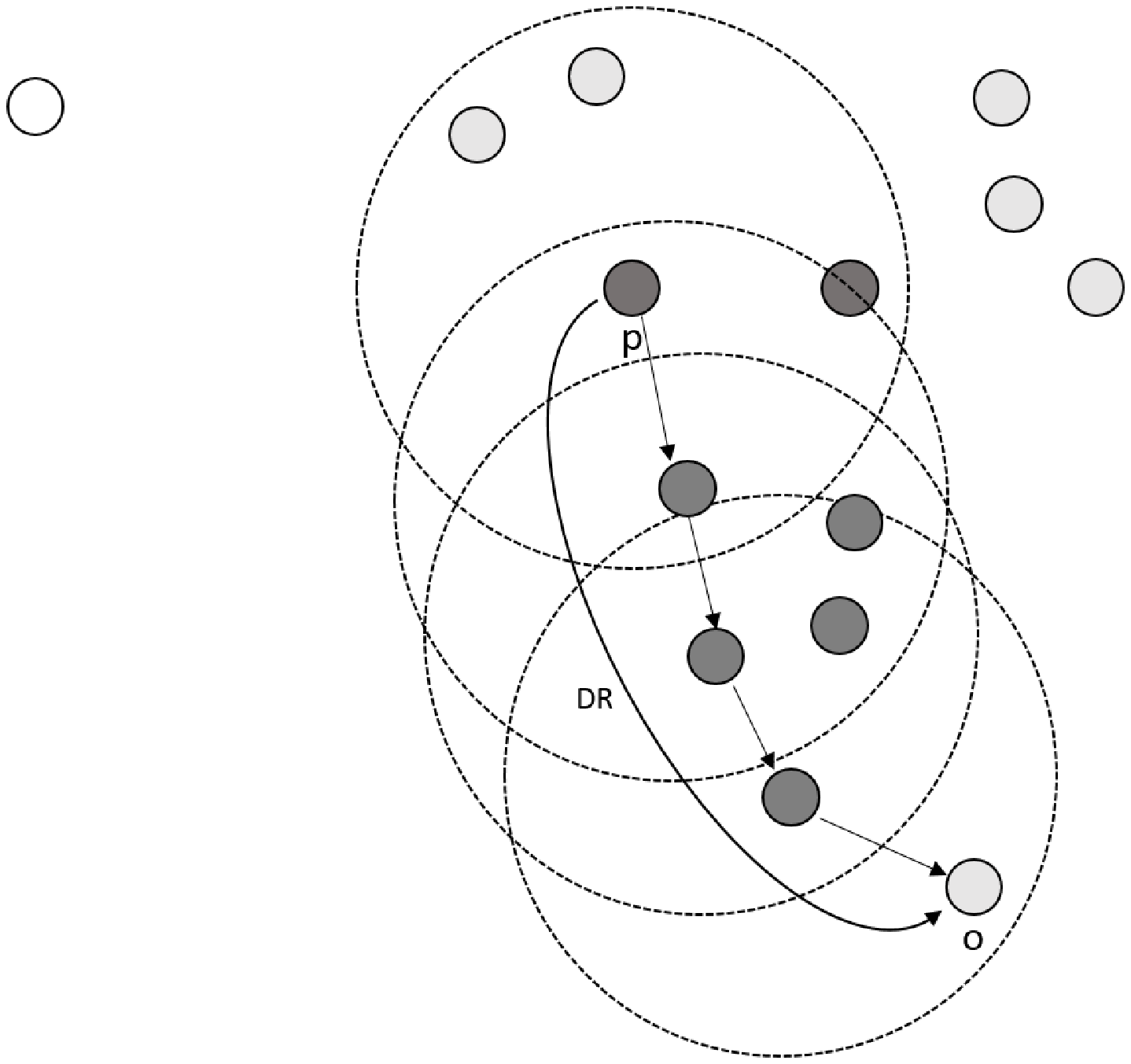}
  \label{fig:sub2}
\end{subfigure}
\caption{The concept of density-reachability with $\eta = 4$. In the left-most picture, $p$ is a core point (dark grey). The middle picture shows that there exists a chain of three directly-density reachable (DDR) connected core points from $p$ to $o$. The $N_{\varepsilon}$ of these connecting points are shown as dotted radius. The right-most pictures shows that $o$ is density-reachable (DR) from $p$.}
\label{fig:dbscan2}
\end{figure}

A point $o$ is \textit{density-reachable} (DR) from a point $p$ if there exists a chain of points which are directly density-reachable from each other. This condition makes DBSCAN flexible for clusters of arbitrary shape due to a chaining effect which is, however, constricted to points which fulfil the two conditions for direct density-reachability. This chaining effect is shown in Figure \ref{fig:dbscan2}.\\

\begin{figure}[h]
\centering
\begin{subfigure}{.5\textwidth}
  \centering
  \includegraphics[width=.9\linewidth]{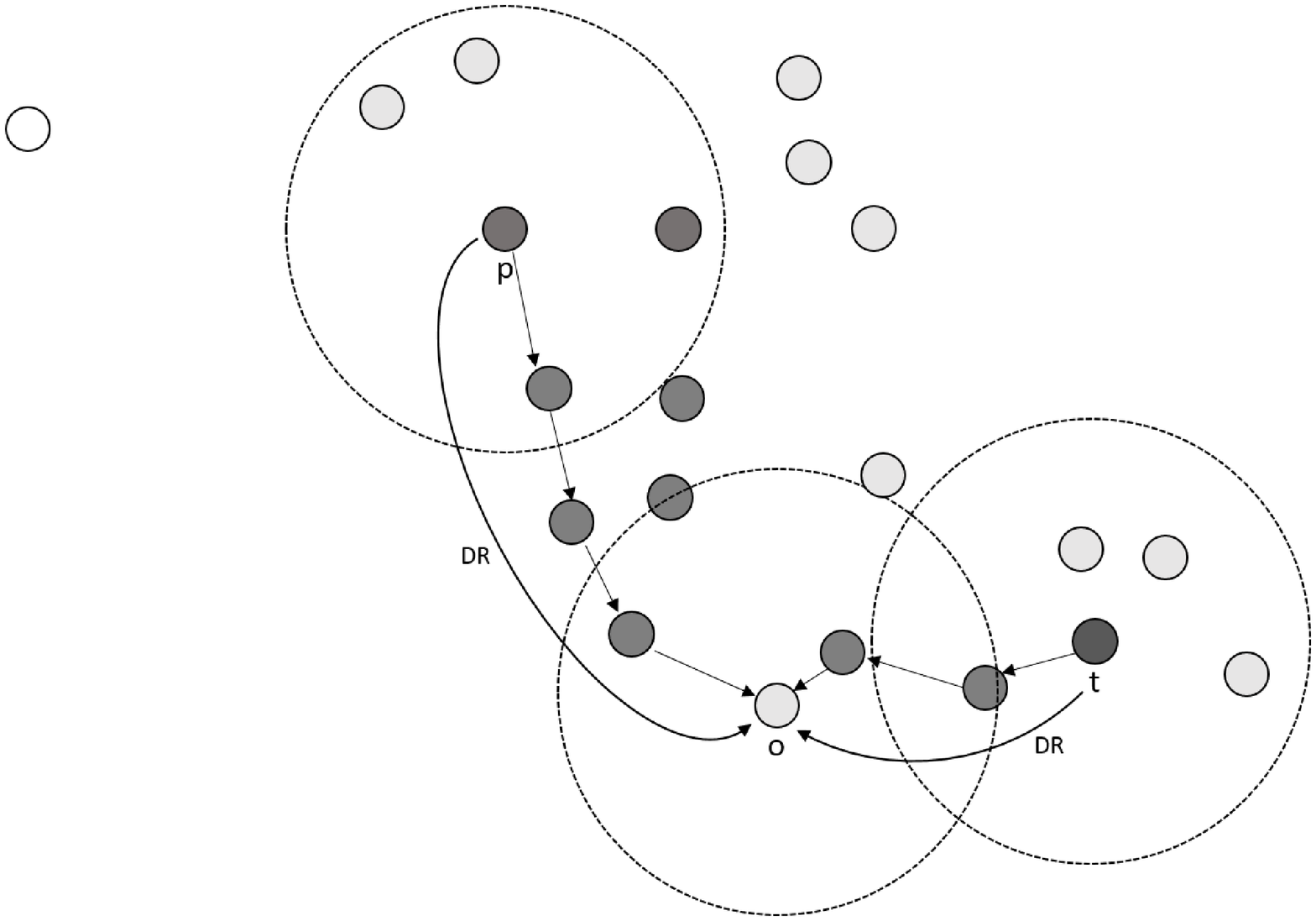}
  \label{fig:sub1}
\end{subfigure}%
\begin{subfigure}{.5\textwidth}
  \centering
  \includegraphics[width=.9\linewidth]{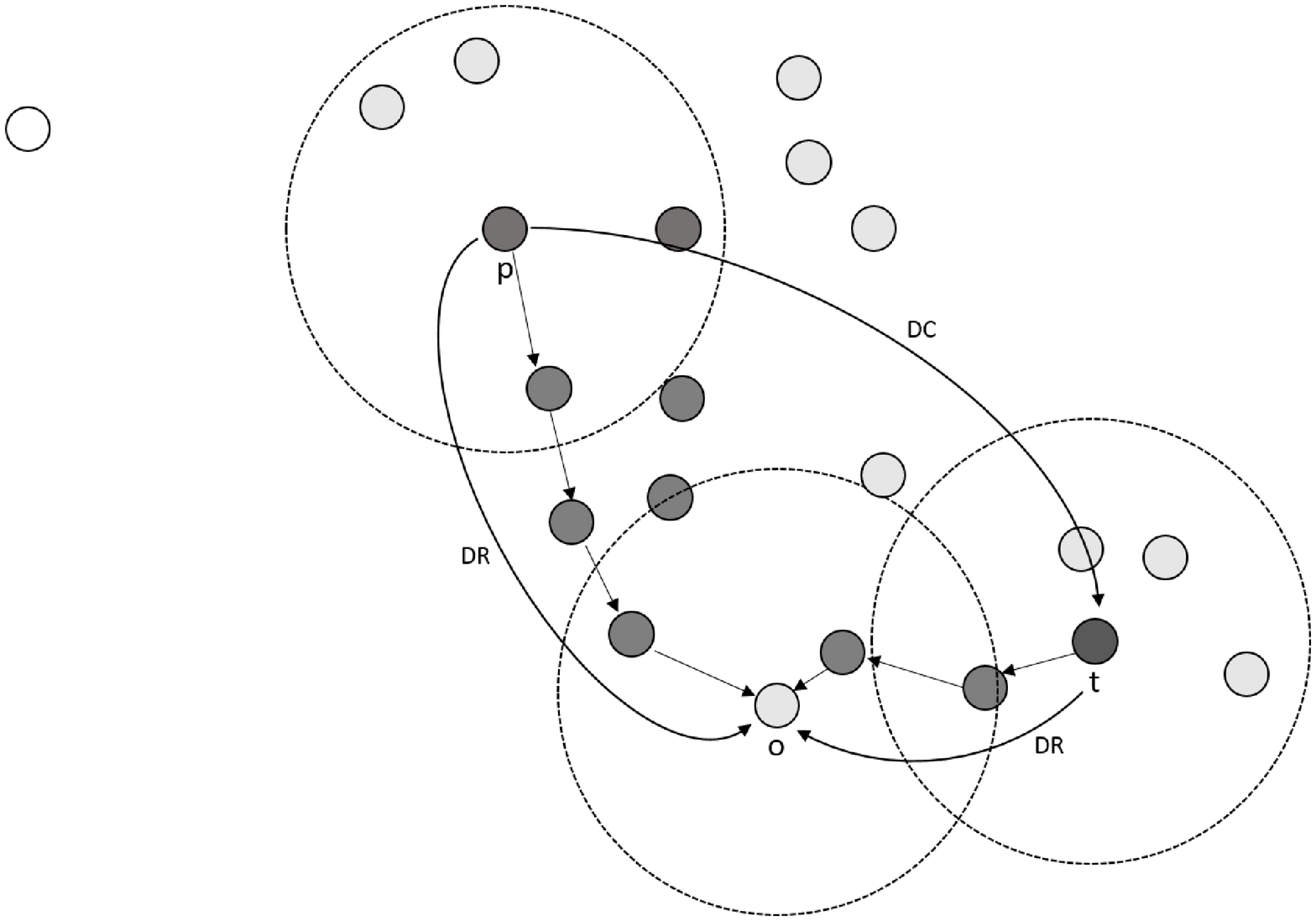}
  \label{fig:sub2}
\end{subfigure}
\caption{The concept of density-connectivity with $\eta = 4$. $t$ is density-connected (DC) to $p$ as there exists a point $o$ which is density-reachable (DR) from both $t$ and $p$ in a similar manner as shown in Figure \ref{fig:dbscan2}.}
\label{fig:dbscan3}
\end{figure}

A point $t$ is \textit{density-connected} (DC) to a point $p$ if there is a point $o$ which is density-reachable from both $t$ and $p$. This concept is shown in Figure \ref{fig:dbscan3}.\\

Based on these definitions, \citeA{ester1996} define a cluster $C$ as a non-empty subset of a database $D$ that fulfils the two conditions

\begin{enumerate}
\item $\forall p,q$: if $p \in C$ and $q$ is density-reachable from $p$ then $q \in C$, and

\item $\forall p,q \in C$: $q$ is density-connected to $p$.

\end{enumerate}

ST-DBSCAN operates in a similar manner as DBSCAN, but in addition to the spatial parameter for determining the $\varepsilon$-neighbourhood, $\varepsilon_{1}$, the algorithm takes an additional parameter $\varepsilon_{2}$, which determines the temporal neighbourhood. In order to be considered a member of a cluster, a data point $q$ has to be in both the spatial and the temporal neighbourhood of point $p$. This extends Equation (\ref{eq:1}) as illustrated in Equation (\ref{eq:2}).

\begin{equation}\label{eq:2}
\begin{split}
N_{\varepsilon_1}(p) = \{q \in D \ | \ \mathrm{dist}(p,q) \leq \varepsilon_1\} \text{  and} \\
N_{\varepsilon_2}(p) = \{q \in D \ | \ \mathrm{dist}(p,q) \leq \varepsilon_2\}.
\end{split} 
\end{equation} 

DBSCAN and ST-DBSCAN both look for neighbouring points for each data point separately, which means that the distance matrices are calculated within the main algorithm body. If one can assume that the spatial distances are stable over time, and that the time series are not updated very frequently, it makes sense to take this matrix calculation outside of the neighbourhood calculation. The two matrices are calculated and stored before running DBSCAN or ST-DBSCAN, after which the algorithm then just has to access the stored information relevant for the respective visited data point.\\

Both DBSCAN and ST-DBSCAN suffer from the drawback of choosing $\varepsilon_1$ for $N_{\varepsilon_1}(p)$. Note that $\varepsilon_{1}$ is global, in the sense that it is constant across the whole region of the study. If the spatial points are distributed with varying densities, as it is, for example, the case in our application in Scotland, then one $\varepsilon_1$ is not be able to capture structures in both dense and sparse areas. If $\varepsilon_{1}$ is chosen too small, it will not be able to identify areas of relatively high density in less dense areas. If it is chosen too large, it will not be able to distinguish between structures in high-density areas. This drawback is the primary motivation for the development of our method. We consider two possible directions to solve this issue. The straightforward approach is to use a distance that is modified according to the density of the area. Alternatively, the global $\varepsilon_{1}$ can be replaced by a local parameter that varies according to the area's density.

\subsection{Kernel density estimation}
\label{methodology:KDE}

In order to make more and less dense areas comparable and, therefore, a global parameter suitable, we employ KDE as a non-parametric way of formulating an area's density. KDE is a density estimation technique which approximates the probability density function from samples. It takes the following form \shortcite{terrell1992, matioli2018}:

\begin{equation}
\widehat{f}(\mathbf{y}) = \frac{1}{nh^{d}} \sum^n_{i=1} K \Big( \frac{\mathbf{x}_i - \mathbf{y}}{h} \Big),
\label{eq:kde}
\end{equation}

for a random sample $\mathbf{x}$ of size $n$, with $\mathbf{x} \in \mathbb{R}^d$, $h$ as a smoothing parameter, and $K$ as the kernel function which satisfies the condition

\begin{equation}
\int_{- \infty}^{\infty} K(x)dx = 1.
\label{eq:kdeKernel}
\end{equation}

KDE is used to estimate the density function from observed data. As a non-parametric approach, it assumes that the data originates from a distribution with the probability density $f$, and then directly uses the data to estimate it. For this, a suitable kernel function is selected, which models the influences of the data points in the data space, with the Gaussian kernel as a commonly used option \cite{hinneburg2007}. The sum of all kernels at a location then provides an estimate of the density function at this location. A Gaussian kernel function is described as

\begin{equation}
K(\mathbf{x}, \mathbf{x}') = \frac{1}{(2 \pi h^2)^{d/2}} e \bigg( - \frac{||\mathbf{x} - \mathbf{x}'||^2}{2h^2} \bigg),
\label{eq:kdeGaussian}
\end{equation}

where $h$ is the bandwidth, $d$ the dimensionality of $\mathbf{x}$, and $||\mathbf{x}|| = \sqrt{\mathbf{x}^T \mathbf{x}}$ denotes the Euclidean norm \cite{sugiyama2015}.\\

Besides selecting the kernel function, there is also the need to select the bandwidth, or smoothing factor, $h$, which affects the smoothness of the surface laid over the points. In other words, it is the influence a data point has on more distant regions , and can be held constant or made variable across the data space \cite{hinneburg2007, terrell1992}. $h$ is often chosen by cross validation or "Scott's Rule' \cite{scott1992}.\\

An advantage of KDE is its continuous nature, as it approximates the density function at all locations of the data space, not just at the locations with existing data points. This makes our approach also suitable for situations in which new data points are added in between existing ones. KDE is often used to identify hotspots which are areas of high point density. \shortciteA{anderson2009} identifies hotspot areas of road accidents with a combination of KDE and $k$-means. They create a density map which is subsequently divided into hotspot cells, with the latter then being clustered using $k$-means. \shortciteA{smith2018} use KDE to create a map of drug-resistant tuberculosis cases in South Africa. \citeA{gerber2014} develops a combined approach of spatio-temporally tagged Twitter data and KDE in order to predict crime in a US city, whereas \citeA{eslinger2017} analyse areas of high numbers of vehicle-bear collisions in Florida. \citeA{matioli2018} introduce ClusterKDE, an algorithm based on univariate KDE combined with optimisation techniques, to iteratively obtain new clusters by minimising a smooth kernel function.

\subsection{ST-DBSCAN with KDE-based local scaling}

Let $\mathbf{x}$ be a set of observations in a two-dimensional space. Using these observations, we estimate the density function with KDE as described in Section \ref{methodology:KDE}. For the selection of the bandwidth $h$, there are a number of optimisation techniques available. We use ``Scott's Rule'' \cite{scott1992}, which calculates $h$ as

\begin{equation}
h = n^{(-1 / (p+4))},
\label{eq:Scott}
\end{equation}

with $n$ as the number of data points and $p$ the number of dimensions. Scott's Rule has previously been shown to perform well compared to other approaches \cite{scott2009}, and is frequently used as a way of determining the optimal bin width, for example by \citeA{bernacchia2011}. Let $d_i$ be the density of point $i$, calculated using Equation (\ref{eq:kdeGaussian}). Let $\bar d$ be the average density of all points in the dataset.\\

Our aim is to re-scale distances in such a way that the result is a distance matrix in which entries for more and less dense areas are comparable. Let $S$ be the spatial distance matrix, then the elements of $S$ are $s_{ij}$ for the distance between points $i$ and $j$, calculated for example using a great-circle distance, which is the shortest distance between two points on a sphere such as the Earth \cite{williams2011}. We then re-scale the distance matrix with weights derived from the density estimates. This means that the resulting scaled distance matrix includes information about the density in the location surrounding an object.\\

A logistic function is chosen for the calculation of the weights from the density estimates. The reasons for this is that the slope of the function can be adjusted by changing a parameter, which makes it possible to change the impact the density estimate has on the weights. Other reasons for choosing a logistic function are its continuous nature and the fact that it can be symmetrically centred around one. The function takes the general form

\begin{equation}
f(x) = \frac{L}{1 + e ^ {-k \cdot (x-x_0)}},
\label{eq:logistic}
\end{equation}

with $x_0$ as the horizontal axis value of the sigmoid's midpoint where the curve changes from convexity to concavity, $L$ as the sigmoid's maximum value, and $k$ as the steepness of the curve.\\

Our aim is to use the function shown in Equation (\ref{eq:logistic}) as a way of ``translating'' density estimates into weights. In order to stretch distances in dense areas and compress them in sparse areas, we use a function that increases or decreases weights while converging to a chosen maximum value. Therefore, in our approach, the curve's midpoint $x_0$ is chosen to be the mean of the density estimates $\mean{d}$ and the curve's maximum $L$ is chosen to be two, thus centring the curve around one on the vertical axis, which gives us for the density $d$ of point $i$

\begin{equation}
f(d_{i}) = \frac{2}{1 + e ^ {-k \cdot (d_{i}-\mean{d})}}.
\label{eq:log}
\end{equation}

This means that if the density estimate for an area is equal to the overall average of all areas, these distances are multiplied with a weight of one, resulting in no modification. The two extreme ends represent the most dense and least dense areas, where the weights are two and zero, respectively. We suggest to weight the distance between two points with a factor of two, the maximum weight, in the highest-density area of the space. We choose the value two as a maximum for symmetry reasons, to account for zero being the minimum and one being the mean weight. The weight of exactly zero is only used to scale the distance between spatially identical points, in which case the distance is already zero. The steepness $k$ of the curve can be adjusted depending on the application.\\ 

The distance matrix is now modified with

\begin{equation}
w_{ij} = s_{ij} \cdot \Big( \frac{f(d_{i}) + f(d_{j})}{2} \Big),
\label{eq:weighting}
\end{equation}


for a point pair $i$ and $j$, and their respective distance $s_{ij}$ in the spatial distance matrix $S$. Equation (\ref{eq:weighting}) calculates the weights for both $i$ and $j$ using Equation (\ref{eq:log}), takes the mean of these two weights, and multiplies this combined weight with the respective distance $s_{ij}$.


\section{Simulation}
\label{simulation}

In order to test the properties of the proposed method, we first apply it to a simulated dataset. The main purpose of this simulation study is to determine whether our approach is able to handle data with varying point densities between different areas, with an improvement in the results when compared to the original algorithm.

\subsection{Set up of the simulations}

\begin{figure}[h]
\centering
\includegraphics[scale=0.22]{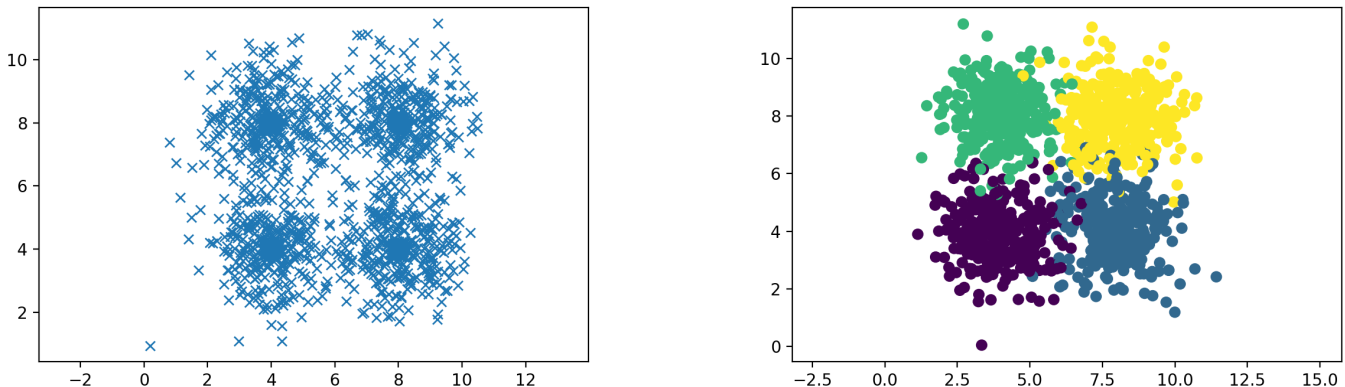}
\caption{2400 data points generated using eight Gaussians consisting of 300 points each. The two axes are random variables $x$ and $y$ from which the points are sampled, with $x,y \in \mathbb{R}$. The left figure shows how two Gaussians, one with $\sigma=1$ and one with $\sigma=0.1$, overlap in such a way that each of the four resulting clusters consists of a high density core and a lower density surrounding. The right figure shows the real cluster memberships by colour, which acts as a way of testing the clustering performance.}
\label{simulationdata}
\end{figure}

For a more straightforward implementation, as we want to highlight the extent to which our proposed distance weighting improves the outcome, we only consider spatial variables. For this purpose, we implement our approach in combination with the original DBSCAN algorithm. An implementation with both spatial and temporal dimensions using ST-DBSCAN can be found in Section \ref{application}.\\
We generate 2400 data points in order to have sufficiently large individual clusters to be detected with our method while being sufficiently small to allow for fast runtimes. They consist of four clusters with 600 randomly Gaussian-distributed points per cluster. More specifically, each cluster consists of 300 points that are randomly sampled from a Gaussian with a higher variance, and 300 points randomly sampled from a Gaussian with a smaller variance and the same mean. The clusters can be seen in Figure \ref{simulationdata}.\\

All four datasets are clustered using the original DBSCAN approach,
the proposed method of DBSCAN with distance weighting, and $k$-means as a benchmark approach.\\
The chosen methods allow us to compare our approach with the original DBSCAN algorithm proposed by \citeA{ester1996} to demonstrate its improved performance under certain conditions. Furthermore, the comparison with $k$-means acts as a benchmark, as $k$-means is considered one of the most common clustering algorithms that also uses distance measures to form clusters (Han et al., 2011, pp.451-454).\\

The results, as displayed in Figure \ref{fig:simulation_result}, are compared based on their ability to identify the more dense clusters while still accounting for any clusters in the less dense areas surrounding them. For assessment purposes, the results are compared regarding their number of outliers and the number of correctly classified points\footnote{All three methods are implemented using R version 3.5.1 and RStudio version 0.99.896. For the original DBSCAN method, we use the R package `dbscan' version 1.1-2, for $k$-means we use the R package `stats' version 3.4.4. Our proposed method uses KDE from the package `ks' version 1.11.0. The code is available from the authors upon request.}.

\subsection{Results of the simulations}

We aim to choose parameters for all three methods using common approaches, comparing them under individually realistic conditions. This results in DBSCAN with $\varepsilon$ = 0.3, chosen according to the nearest neighbour plot as proposed by \citeA{ester1996}, and $\eta$ = 6. For the proposed adapted DBSCAN, we choose $\varepsilon$ = 0.3 and $\eta$ = 6 based on the same principle. For $k$-means, we choose $k=4$ based on both the known real number of clusters and the within-clusters sum of squares (`elbow method').\\

\begin{figure}[!htb]
\centering
\includegraphics[scale=0.15]{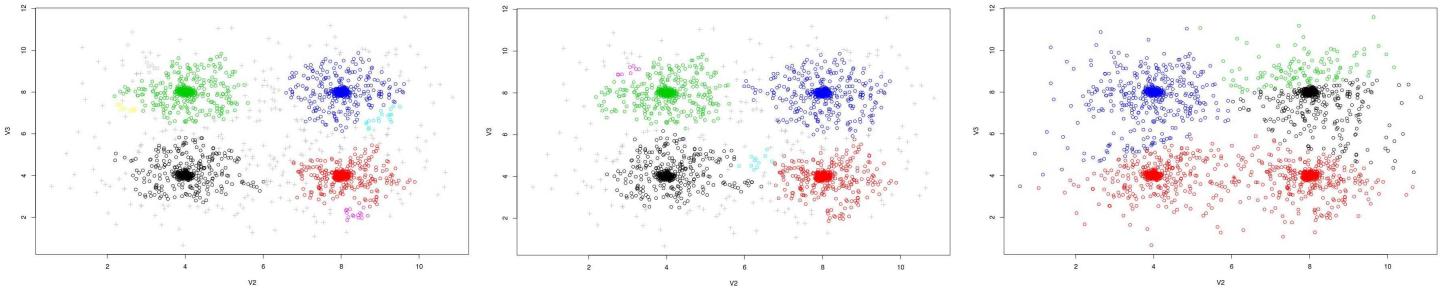}
\caption{Clustering results for three tested algorithms using the generated simulation data. The axes refer to the random variables $x$ and $y$ from which the Gaussians were sampled as shown in Figure \ref{simulationdata}. The left figure shows the result for the original DBSCAN algorithm. The middle figure shows the result for the proposed adapted DBSCAN algorithm. The right figure shows the result for $k$-means. Both the original DBSCAN and the proposed method correctly separate the four clusters. $K$-means combines the two bottom clusters into one (red coloured). }
\label{fig:simulation_result}
\end{figure}

The initial assessment is a general visual comparison of the results as shown in Figure \ref{fig:simulation_result}. As the data is generated to represent four clusters, we are able to assess whether the three methods correctly identify these four groups.\\

\begin{table}[h]
\begin{center}
\begin{tabular}{| c | c | c | c |}
\hline
 & Original DBSCAN & Proposed method & $k$-means \\
\hline
Number of outliers & 256 & 183 & NA\\
\hline
Correctly clustered elements & 2092 & 2175 & 1189\\
\hline
Incorrectly clustered elements & 308 & 225 & 1211\\
\hline 
\end{tabular}
\caption{Number of outliers and number of correctly and incorrectly clustered data points for each algorithm when tested on the generated data set. The proposed method has a higher number of correctly clustered elements than both the original DBSCAN and $k$-means, and a lower number of outliers than the original DBSCAN.}
\label{table:simulation}
\end{center}
\end{table}

Next, we compare the number of outliers, as well as the number of correctly clustered objects as shown in Table \ref{table:simulation}. As expected, $k$-means does not perform well on the simulated data set, combining the two lower clusters into one and splitting the top right cluster into two. The reason for this can likely be found in the random initial seeds, which strongly influence the clustering outcome. The number of outliers is not applicable (NA) for this method, as $k$-means assigns each data point to a cluster without marking outliers.\\
The results for DBSCAN and the proposed method show an improvement for the latter regarding the number of correctly and incorrectly clustered objects. Looking closely at the graphical outputs shows that the proposed method is able to capture more of the less densely-spread data points surrounding the denser areas. Due to its ability to adapt to changing environments, the algorithm's result has a lower number of outliers because the search radius parameter varies with changing point densities. This makes the method especially valuable when outliers are driven due to environmental influences such as more rural or remote areas, and the researcher wishes to account for this fact.

\section{Empirical application}
\label{application}

Our proposed method is tested for its applicability by analysing prescription data on GPs from the Scottish NHS. The chosen data was collected over a two-year period from October 2015 to September 2017. There are four main objectives for this application. The first is to improve the general understanding of GP behaviour over time as well as differences between GPs in different locations across Scotland. As a further development, the usage of clustering also enables the organisation to generate representative samples of GPs. Sampling from the generated clusters means that the existing underlying structures can be replicated. Additional analysis of GPs can then be conducted using the samples instead of the whole dataset, improving the speed of analysis significantly and allowing for real-time analysis and visualisation tools. Lastly, the developed software tool, which was built based on our methodology, makes our approach accessible for users from different areas to improve their data understanding in an interactive and real-time manner.

\subsection{Data source and description}

The data sources are NHS Scotland publications and the Open Data Platform of their Information Services Division (ISD)\footnote{https://www.opendata.nhs.scot/}. Using open data has multiple advantages, such as being readily accessible via the ISD's website which also makes our research easy to replicate.\\

\begin{figure}[h]
\centering
\includegraphics[scale=0.6]{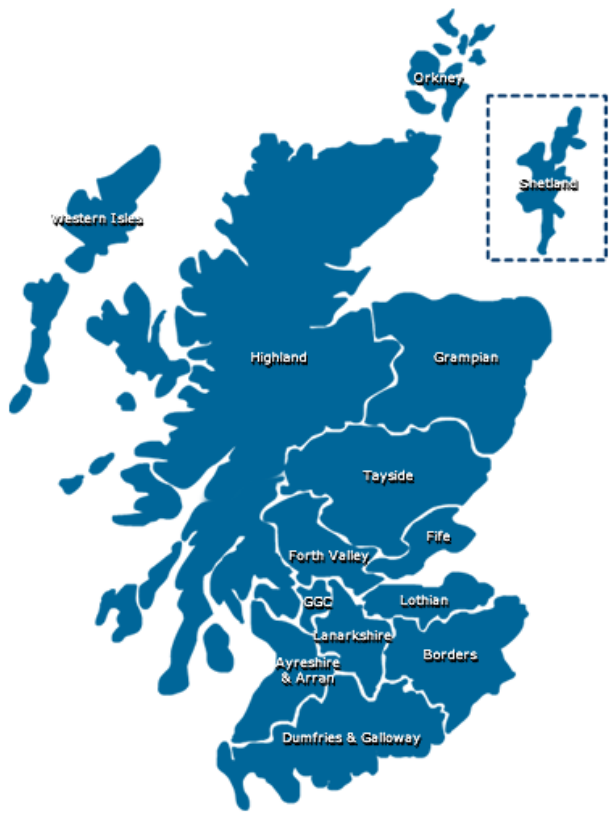}
\caption{Healthboard areas of NHS Scotland. The figure shows a map of Scotland and the fourteen territorial healthboards into which it is divided. Source: NHS Scotland (2013).}
\label{fig:healthboards}
\end{figure}

NHS Scotland covers 14 territorial healthboards of Scotland as shown in Figure \ref{fig:healthboards} and 4,994 GPs as of September 2018 \cite{nhsisd2018}. The number of GPs varies across the different healthboards, as seen in Table \ref{table:healthboards}.

\begin{table}[h!]
\begin{center}
\begin{tabular}{|c|c|c|c|}
\hline
\textbf{Healthboard} & \textbf{Number of GPs} & \textbf{Healthboard} & \textbf{Number of GPs}\\
\hline
Ayrshire \& Arran & 338 & Borders & 126\\
\hline
Dumfries \& Galloway & 120 & Fife & 280\\
\hline
Forth Valley & 270 & Grampian & 533\\
\hline
Greater Glasgow and Clyde & 1,076 & Highland & 398\\
\hline
Lanarkshire & 441 & Lothian & 941\\
\hline
Orkney & 36 & Shetland & 30\\
\hline
Tayside & 395 & Western Isles & 32\\
\hline
\end{tabular}
\caption{Number of GPs for each territorial healthboard in Scotland. Each of the fourteen healthboards is listed with their respective GP `headcount'. The number of GPs varies greatly between the different healthboards, from as low as 30 to as high as 1076.}
\label{table:healthboards}
\end{center}
\end{table}

The prescription data from NHS ISD is merged with additional open data containing the GPs' locations given by their postcode and corresponding latitude and longitude. We filter the data table for one specific drug, Amoxicillin, which is an antibiotic that is used to treat bacterial infections. The time series for each GP is then calculated by aggregating the prescription instances for different types of the drug on a monthly basis. Based on a two-year period from October 2015 to September 2017 with monthly data, this gives us a time series of length 24 for each GP. Additional variables are a unique identifier (practice code), post code, latitude, and longitude, which account for four more columns. There are 1081 rows in total, which includes all GPs who prescribed any type of Amoxicillin over the period of two years. Due to missing values in their coordinates, a number of rows had to be removed, resulting in a 980$\times$28 matrix. 

\begin{table}[h!]
\begin{center}
\begin{tabular}{| c | c | c | c | c | c | c | c |}
\hline
v1 & v2 & v3 &  & v26 & v27 & v28 \\
\hline
Practice code & Postcode & 201510 & ... & 201709 & Latitude & Longitude\\
\hline
\end{tabular}
\caption{Variables in the aggregated data matrix after filtering. Each GP practice which prescribed Amoxicillin between October 2015 and September 2017 is represented by a table row using a unique practice code as an identifier. The table lists their postcode and the corresponding latitude and longitude values, as well as their individual time series of summed up Amoxicillin prescriptions per month.}
\end{center}
\end{table}

\begin{figure}[!h]
\centering
\includegraphics[scale=0.6]{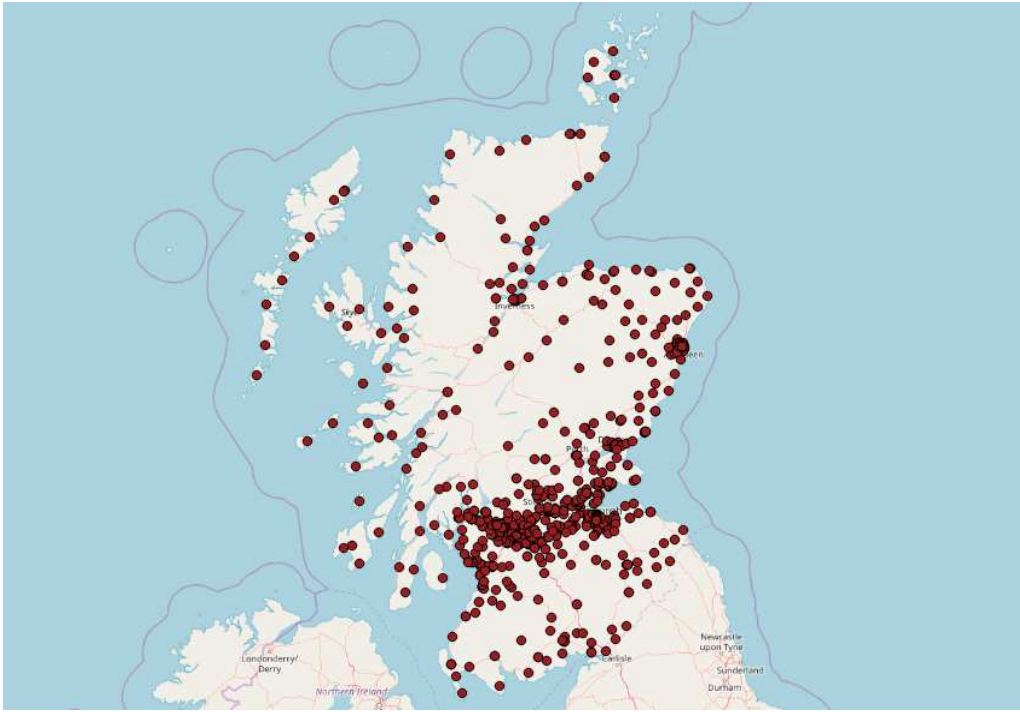}
\caption{Map of Scotland with relevant GP locations. The map shows the location of each GP in Scotland who prescribed Amoxicillin between October 2015 and September 2017. The more densely populated South of the country as well as the more sparsely populated North and North-West are clearly visible.}
\label{ScotlandMap}
\end{figure}

When plotting the locations of the 980 analysed GPs using their latitude and longitude, as shown in Figure \ref{ScotlandMap}, it becomes clear the distribution of their locations is not even across the country. This is likely to be the case due to the uneven population distribution in Scotland. The ``belt'' of more densely populated areas towards the South is clearly visible, as are the much less populated northern regions.

\begin{figure}[h!]
\centering
\includegraphics[scale=0.5]{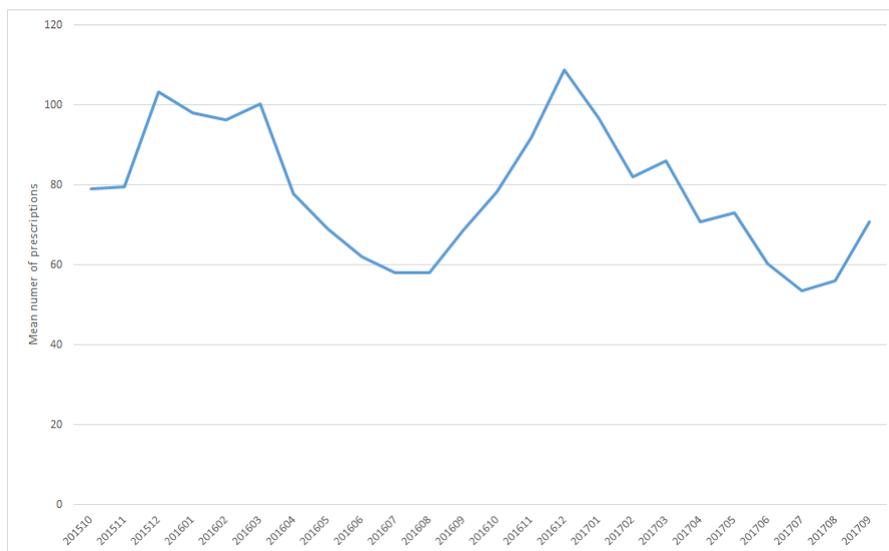}
\caption{Amoxicillin prescription counts in Scotland from October 2015 to September 2017. The figure shows the average sum of prescriptions per month for all analysed GPs. The line shows seasonality, with higher prescription volumes during the winter months compared to the summer.}
\label{meantimeseries}
\end{figure}

Figure \ref{meantimeseries} shows the mean sum of prescriptions over all GPs per month. As can be seen, more Amoxicillin is prescribed during the late autumn and winter months (October to January) compared to the summer months (May to August).

\subsection{Clustering process and results}

We use two algorithms to cluster our data. In a first step, we use the original ST-DBSCAN algorithm. The spatial distance matrix is calculated using the `Meeus' great-circle distance from the R package `geosphere' \cite{hijmans2017}. The temporal dissimilarity between the time series is calculated using the DTW function `DTWDistance' from the R package `TSDist' \shortcite{mori2016}. These two dissimilarity matrices are then used by the algorithm to define $\varepsilon$-neighbourhoods as described in Section \ref{method:dbscan}.\\

The time series are clustered by calculating the temporal dissimilarity matrices using DTW. Due to the expected seasonal variations in the prescription levels of Amoxicillin, DTW was chosen as it is able to detect similarity between time series where peaks in prescription volume occur shifted. This seasonality is shown in Figure \ref{meantimeseries}. The ST-DBSCAN algorithm calculates the dissimilarity matrix outside of the main algorithm body and refers back to it during the $N_{\varepsilon_2}$ search. $\varepsilon_2=1000$ is chosen using the nearest neighbour distance plot as proposed by \citeA{ester1996}.\\

\begin{figure}[h!]
\centering
\includegraphics[width=0.9\textwidth]{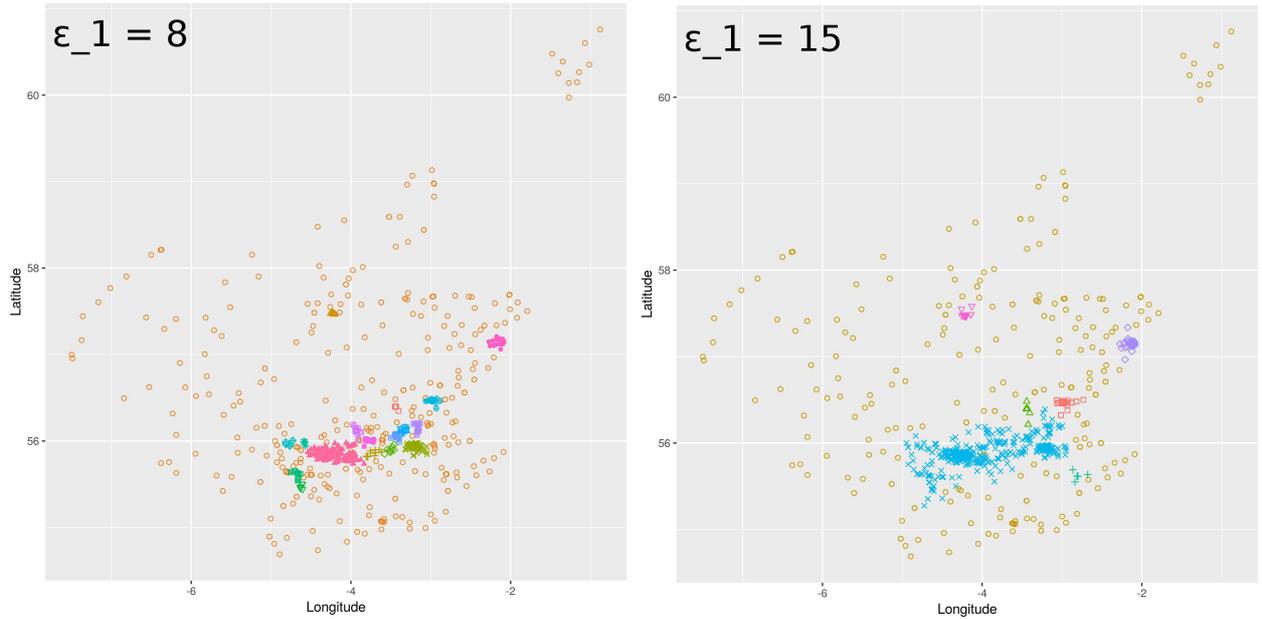}
\caption{Clustering results for the original ST-DBSCAN algorithm. The upper figure shows the results with $\varepsilon_{1}=8$, which means an 8km radius. Each colour represents one cluster, the orange data points represent outliers. With the smaller radius, the algorithm is able to cluster mostly dense areas into several small clusters, especially in the ``belt'' area which consists of Edinburgh and Glasgow. The lower figure shows results with $\varepsilon_{1}=15$, a 15km radius. With the larger radius, the algorithm is not able to cluster dense areas. Instead, it combines them into one large cluster marked in blue.}
\label{ExpOriginalDBSCAN}
\end{figure}

The uneven spatial data distribution does, however, leads to a problem. The global spatial parameter $\varepsilon_{1}$ is not able to capture small-scale clusters in the South without oversmoothing them. If chosen too small, the algorithm will form many (usually around eleven) very small and dense clusters, with many points within less dense areas marked as outliers. Alternatively, the algorithm will select all points in the denser areas to belong to the same cluster when the chosen parameter is too large, as seen in Figure \ref{ExpOriginalDBSCAN}.\\

The overall goal of our proposed approach is for the algorithm to not overemphasise clusters in dense regions such as cities. This thought leads to the idea of scaling the distances based on their surrounding point density. For this, we use the process described in Section \ref{methodology}. We calculate the density estimates for all regions based on the amount of data points present in each area. Using these estimates as a way of determining how dense or sparse a geographic region is, we derive weights with which we multiply our spatial distance matrix. The updated distance matrix is then used as an input for the original ST-DBSCAN algorithm.\\

\begin{figure}[h!]
\centering
\includegraphics[width=0.8\textwidth]{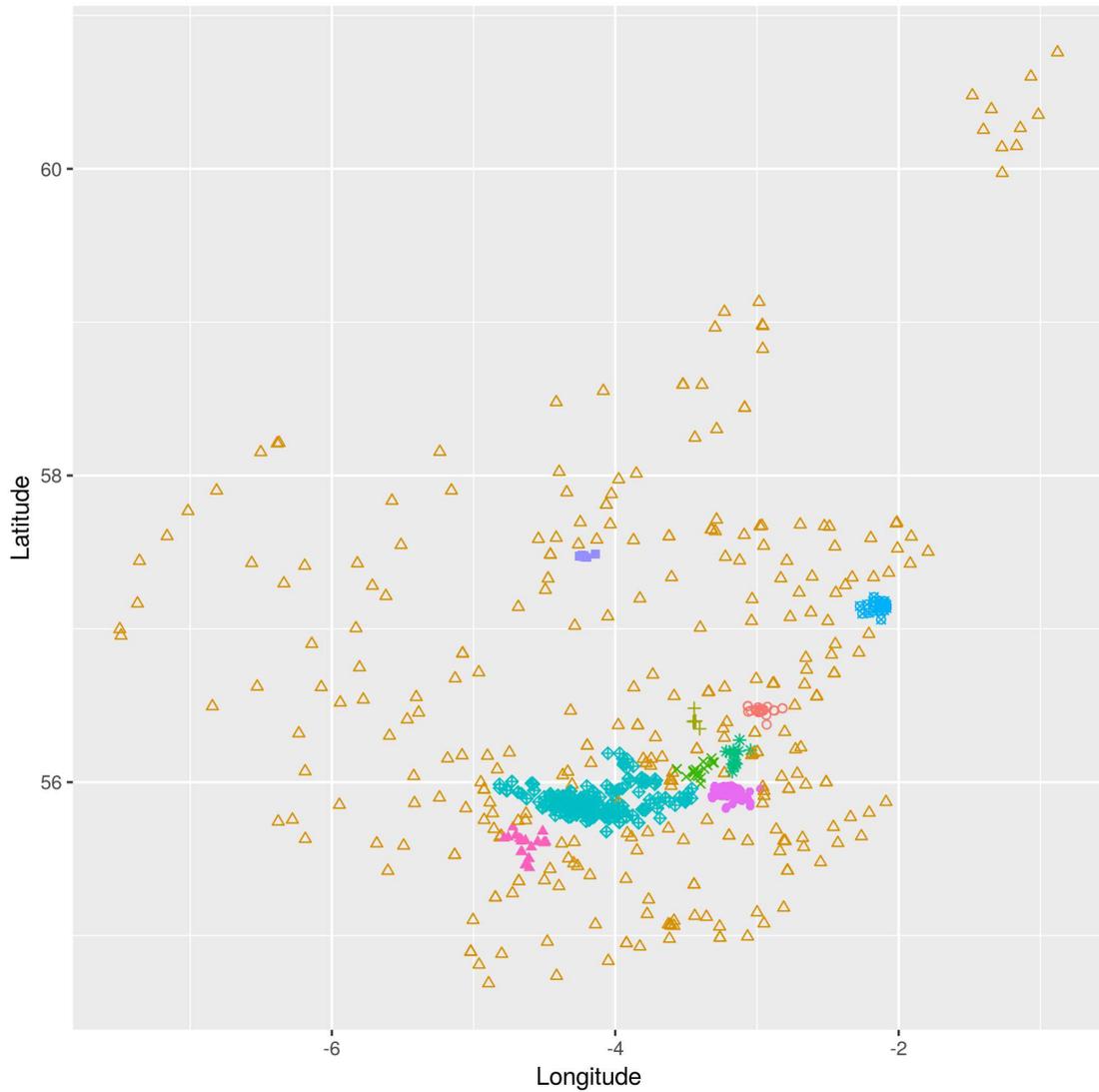}
\caption{Adapted approach with $\varepsilon_{1}=8$, the same $\varepsilon_{1}$ as the upper panel of Figure \ref{ExpOriginalDBSCAN}. Each cluster is represented by a different colour. Our approach is able to cluster the dense area and combine some of the smaller clusters in the latter into two larger clusters, marked in turquoise and green.}
\label{ExpAdaptDBSCAN}
\end{figure}

The results for our implementation show a more realistic clustering result, which is more useful for organisations working on such problems, as seen in Figure \ref{ExpAdaptDBSCAN}. For both the original ST-DSBCAN algorithm and our proposed method, we choose $\varepsilon_{1}$ = 8, $\varepsilon_{2}$ = 1000, and $\eta$ = 7 according to nearest neighbour distance plots as proposed by \citeA{ester1996}. With the same parameters, our adapted ST-DBSCAN method typically captures around six clusters in the belt area compared to the eleven small ones with the original method. Overall, our approach identifies a smaller number of nine clusters, compared to 17 for the original ST-DBSCAN algorithm, as it is able to combine some smaller dense areas with their surrounding points. 

\begin{figure}[h!]
\centering
\includegraphics[scale=0.25]{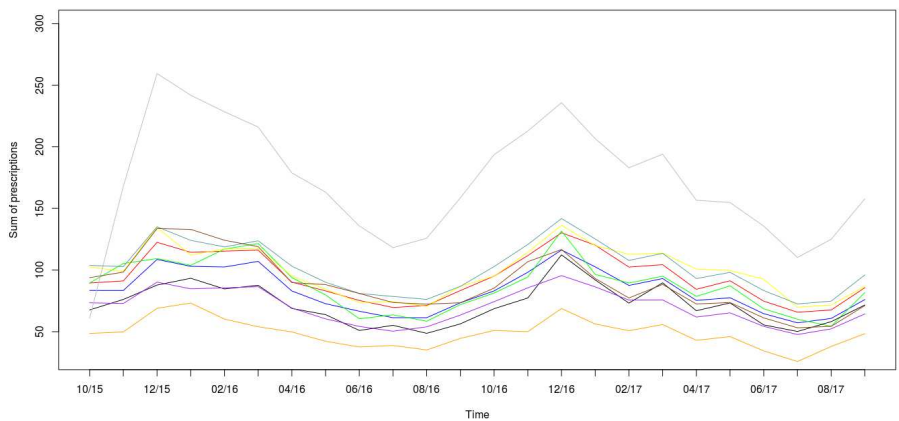}
\caption{Average Amoxicillin prescription volume over time in each of the nine identified clusters. The clusters show a very similar seasonality pattern which only differs in volume and not, for example, in the point in time at which a peak occurs.}
\label{clustertimeseries}
\end{figure}

Our proposed approach is able to highlight extreme behaviour such as particularly high or low prescription volumes of single or groups of GPs. This makes the results especially interesting for policy makers and the NHS as a whole, as unusually high prescription levels of a specific drug could be an indicator for an underlying problem in this region. Figure \ref{ExpAdaptDBSCAN} shows, however, that many points are still classified as outliers. This can be explained due to the chosen variables, meaning prescription volume over time and location, not being informative enough to fully cluster GPs. Additional variables such as socio-economic factors can potentially add more useful information to this problem. Figure \ref{clustertimeseries}, which shows the average prescription amounts in each cluster and their seasonal changes over the analysed time period, illustrates this issue. Each cluster has a very similar time series, which mostly differs in total prescription volume and not, as one would expect, in the point in time where peaks occur. This exemplifies the need for further research, for example by analysing a different drug or adding additional factors.

\section{Conclusion}
\label{conclusion}

The well-known DBSCAN algorithm \cite{ester1996} and its spatio-temporal adaptation ST-DBSCAN \cite{birant2007} are flexible and efficient ways of clustering spatial and spatio-temporal data \shortcite{catlett2019, chimwayi2018, gomide2011}. Due to its nature, DBSCAN suffers from a weakness when presented with clusters of varying densities. This is frequently the case when using DBSCAN for real-world spatial data. Previous research has emphasised this issue and tried to solve it using nearest neighbour approaches \cite{pei2009, biccici2007, zelnik2005, ertoez2003}. These approaches do, however, not result in a continuous representation and handling of the varying densities.\\
When clustering time series, the choice of dissimilarity measure is crucial. In order to cluster shifted time series, elastic measures such as DTW offer a suitable solution and perform better than step-wise methods. To the best of our knowledge, ST-DSBCAN has not been used in combination with DTW before.\\

In this paper, we present a method of scaling distances using continuous density-based distance weights. Our objective is to introduce a method of manipulating a spatial dissimilarity or distance matrix in such a way that it reflects the density of points in a given area. For this purpose, we propose and implement a way of deriving weights from the density estimates acquired through KDE, using a logistic function. Our proposed method first estimates the density in each spatial region, which is then used to weight the distances between data points depending on how dense or sparse their respective areas are. In denser areas, a higher weighting factor leads to the stretching of a distance, whereas lower weights for less dense areas lead to the compression of a distance.\\ 

We demonstrate the applicability of this approach for clustering with density-based methods such as DBSCAN, especially in cases of uneven data point densities in which a global distance parameter might not be able to capture group structures in regions of varying density. For this purpose, we show how our approach works well in both experiments with simulated data and an application case of NHS Scotland prescription data. The second application case also establishes that our approach works well in combination with using DTW for the temporal dimension. In addition, our method is very flexible and can be used in connection with any clustering algorithm that uses a distance matrix as an input.\\

We see the methodological contribution of our work in the use of distance scaling based on local densities using KDE. The inability of DBSCAN to handle varying point densities has been acknowledged by some of the algorithm's original authors in \citeA{kriegel2011}. Previous attempts to tackle this issue focus on nearest neighbour approaches \cite{pei2009, biccici2007, zelnik2005, ertoez2003} or assigning density factors to clusters \cite{birant2007}. As for the reason behind these developments, \citeA{biccici2007} comment on their choice of a neighbour-based density estimate instead of a Parzen window approach by highlighting the difficulty of choosing the right window size. This issue has, however, been addressed extensively in the literature covering KDE. Our approach also offers added benefits due to the use of a continuous density map, which is not only useful for measuring the density for any location or new point, but also has the advantage of good visualisation possibilities. In contrast to nearest neighbour approaches, our proposed method does not require a specification of the number of neighbours considered to assess an area's density. We thus see our algorithm as an improvement, especially in terms of interpretability and visualisation. Its flexibility means that it can be applied to any case in which data points are spatially clustered with an algorithm that uses a distance matrix as an input. \\

Furthermore, with regard to the application case and practical contribution, our research offers actionable insights into drug prescription behaviour. Our findings demonstrate how clustering can be used to group GPs according to their spatial location and behaviour over time. Our proposed method is able to highlight extreme behaviour such as GPs deviating from peers in their group. This has been acknowledged as an important issue by both the NHS and, in the special case of antibiotics, the British government \cite{healthcare2018, department2016}, showcasing the increasing awareness of policy makers.\\ 

We see the potential for future research in exploring how the smoothing factor in KDE affects the results and can be adjusted depending on the application case. When interpreted in the context of our approach, the smoothing factor accounts for the amount of impact a group of data points has on their surroundings, or how far their ``density-impact'' reaches spatially. From an application point of view, this should be considered by researchers on a case-by-case basis. \\

In addition, our application to prescription data has implications for the analysis of prescription behaviour in different areas and over time. While promising, our findings also provide an example for the issue of trying to explain complex dynamics using only two variables. The fact that a relatively common antibiotic has been chosen as the drug to be analysed might also contribute to the issue of relatively little spatial fluctuation. Further research could focus on additional explanatory variables which can shed more light on this behaviour, for example socioeconomic factors, as additional dimensions besides time and space. Research similar to this proposal is, for example, introduced by \citeA{kjaerulff2016}. This will help to further expand the understanding of prescription behaviour for different medications and in different regions, offering valuable insights for organisation such as the NHS and policy makers alike.


\section*{Acknowledgements}
The authors thank Wallscope for the support provided with the data access.\\
Funding: This work was supported by The Data Lab [project registration number Reg-17669].

\setstretch{1.47}
\bibliography{mybib}

\begin{thebibliography}{}

\bibitem [\protect \citeauthoryear {%
Aghabozorgi%
, Shirkhorshidi%
\BCBL {}\ \BBA {} Wah%
}{%
Aghabozorgi%
\ \protect \BOthers {.}}{%
{\protect \APACyear {2015}}%
}]{%
aghabozorgi2015}
\APACinsertmetastar {%
aghabozorgi2015}%
\begin{APACrefauthors}%
Aghabozorgi, S.%
, Shirkhorshidi, A\BPBI S.%
\BCBL {}\ \BBA {} Wah, T\BPBI Y.%
\end{APACrefauthors}%
\unskip\
\newblock
\APACrefYearMonthDay{2015}{}{}.
\newblock
{\BBOQ}\APACrefatitle {Time-series clustering--A decade review} {Time-series
  clustering--a decade review}.{\BBCQ}
\newblock
\APACjournalVolNumPages{Information Systems}{53}{}{16--38}.
\newblock
\begin{APACrefDOI} \doi{10.1016/j.is.2015.04.007} \end{APACrefDOI}
\PrintBackRefs{\CurrentBib}

\bibitem [\protect \citeauthoryear {%
Anbaroglu%
, Heydecker%
\BCBL {}\ \BBA {} Cheng%
}{%
Anbaroglu%
\ \protect \BOthers {.}}{%
{\protect \APACyear {2014}}%
}]{%
anbaroglu2014}
\APACinsertmetastar {%
anbaroglu2014}%
\begin{APACrefauthors}%
Anbaroglu, B.%
, Heydecker, B.%
\BCBL {}\ \BBA {} Cheng, T.%
\end{APACrefauthors}%
\unskip\
\newblock
\APACrefYearMonthDay{2014}{}{}.
\newblock
{\BBOQ}\APACrefatitle {Spatio-temporal clustering for non-recurrent traffic
  congestion detection on urban road networks} {Spatio-temporal clustering for
  non-recurrent traffic congestion detection on urban road networks}.{\BBCQ}
\newblock
\APACjournalVolNumPages{Transportation Research Part C: Emerging
  Technologies}{48}{}{47--65}.
\newblock
\begin{APACrefDOI} \doi{10.1016/j.trc.2014.08.002} \end{APACrefDOI}
\PrintBackRefs{\CurrentBib}

\bibitem [\protect \citeauthoryear {%
C.~Anderson%
, Lee%
\BCBL {}\ \BBA {} Dean%
}{%
C.~Anderson%
\ \protect \BOthers {.}}{%
{\protect \APACyear {2016}}%
}]{%
anderson2016}
\APACinsertmetastar {%
anderson2016}%
\begin{APACrefauthors}%
Anderson, C.%
, Lee, D.%
\BCBL {}\ \BBA {} Dean, N.%
\end{APACrefauthors}%
\unskip\
\newblock
\APACrefYearMonthDay{2016}{}{}.
\newblock
{\BBOQ}\APACrefatitle {Bayesian cluster detection via adjacency modelling}
  {Bayesian cluster detection via adjacency modelling}.{\BBCQ}
\newblock
\APACjournalVolNumPages{Spatial and spatio-temporal
  epidemiology}{16}{}{11--20}.
\newblock
\begin{APACrefDOI} \doi{10.1016/j.sste.2015.11.005} \end{APACrefDOI}
\PrintBackRefs{\CurrentBib}

\bibitem [\protect \citeauthoryear {%
T.~Anderson%
}{%
T.~Anderson%
}{%
{\protect \APACyear {2009}}%
}]{%
anderson2009}
\APACinsertmetastar {%
anderson2009}%
\begin{APACrefauthors}%
Anderson, T.%
\end{APACrefauthors}%
\unskip\
\newblock
\APACrefYearMonthDay{2009}{}{}.
\newblock
{\BBOQ}\APACrefatitle {Kernel density estimation and K-means clustering to
  profile road accident hotspots} {Kernel density estimation and k-means
  clustering to profile road accident hotspots}.{\BBCQ}
\newblock
\APACjournalVolNumPages{Accident Analysis \& Prevention}{41}{3}{359--364}.
\newblock
\begin{APACrefDOI} \doi{10.1016/j.aap.2008.12.014} \end{APACrefDOI}
\PrintBackRefs{\CurrentBib}

\bibitem [\protect \citeauthoryear {%
Ankerst%
, Breunig%
, Kriegel%
\BCBL {}\ \BBA {} Sander%
}{%
Ankerst%
\ \protect \BOthers {.}}{%
{\protect \APACyear {1999}}%
}]{%
ankerst1999}
\APACinsertmetastar {%
ankerst1999}%
\begin{APACrefauthors}%
Ankerst, M.%
, Breunig, M\BPBI M.%
, Kriegel, H\BHBI P.%
\BCBL {}\ \BBA {} Sander, J.%
\end{APACrefauthors}%
\unskip\
\newblock
\APACrefYearMonthDay{1999}{}{}.
\newblock
{\BBOQ}\APACrefatitle {OPTICS: ordering points to identify the clustering
  structure} {Optics: ordering points to identify the clustering
  structure}.{\BBCQ}
\newblock
\BIn{} \APACrefbtitle {ACM Sigmod record} {Acm sigmod record}\ (\BVOL~28,
  \BPGS\ 49--60).
\PrintBackRefs{\CurrentBib}

\bibitem [\protect \citeauthoryear {%
Benati%
, Puerto%
\BCBL {}\ \BBA {} Rodr{\'\i}guez-Ch{\'\i}a%
}{%
Benati%
\ \protect \BOthers {.}}{%
{\protect \APACyear {2017}}%
}]{%
benati2017}
\APACinsertmetastar {%
benati2017}%
\begin{APACrefauthors}%
Benati, S.%
, Puerto, J.%
\BCBL {}\ \BBA {} Rodr{\'\i}guez-Ch{\'\i}a, A\BPBI M.%
\end{APACrefauthors}%
\unskip\
\newblock
\APACrefYearMonthDay{2017}{}{}.
\newblock
{\BBOQ}\APACrefatitle {Clustering data that are graph connected} {Clustering
  data that are graph connected}.{\BBCQ}
\newblock
\APACjournalVolNumPages{European Journal of Operational
  Research}{261}{1}{43--53}.
\newblock
\begin{APACrefDOI} \doi{10.1016/j.ejor.2017.02.009} \end{APACrefDOI}
\PrintBackRefs{\CurrentBib}

\bibitem [\protect \citeauthoryear {%
Bernacchia%
\ \BBA {} Pigolotti%
}{%
Bernacchia%
\ \BBA {} Pigolotti%
}{%
{\protect \APACyear {2011}}%
}]{%
bernacchia2011}
\APACinsertmetastar {%
bernacchia2011}%
\begin{APACrefauthors}%
Bernacchia, A.%
\BCBT {}\ \BBA {} Pigolotti, S.%
\end{APACrefauthors}%
\unskip\
\newblock
\APACrefYearMonthDay{2011}{}{}.
\newblock
{\BBOQ}\APACrefatitle {Self-consistent method for density estimation}
  {Self-consistent method for density estimation}.{\BBCQ}
\newblock
\APACjournalVolNumPages{Journal of the Royal Statistical Society: Series B
  (Statistical Methodology)}{73}{3}{407--422}.
\newblock
\begin{APACrefDOI} \doi{10.1111/j.1467-9868.2011.00772.x} \end{APACrefDOI}
\PrintBackRefs{\CurrentBib}

\bibitem [\protect \citeauthoryear {%
Bi{\c{c}}ici%
\ \BBA {} Yuret%
}{%
Bi{\c{c}}ici%
\ \BBA {} Yuret%
}{%
{\protect \APACyear {2007}}%
}]{%
biccici2007}
\APACinsertmetastar {%
biccici2007}%
\begin{APACrefauthors}%
Bi{\c{c}}ici, E.%
\BCBT {}\ \BBA {} Yuret, D.%
\end{APACrefauthors}%
\unskip\
\newblock
\APACrefYearMonthDay{2007}{}{}.
\newblock
{\BBOQ}\APACrefatitle {Locally scaled density based clustering} {Locally scaled
  density based clustering}.{\BBCQ}
\newblock
\BIn{} \APACrefbtitle {International Conference on Adaptive and Natural
  Computing Algorithms} {International conference on adaptive and natural
  computing algorithms}\ (\BPGS\ 739--748).
\newblock
\begin{APACrefDOI} \doi{10.1007/978-3-540-71618-1_82} \end{APACrefDOI}
\PrintBackRefs{\CurrentBib}

\bibitem [\protect \citeauthoryear {%
Bie%
, Mehmood%
, Ruan%
, Sun%
\BCBL {}\ \BBA {} Dawood%
}{%
Bie%
\ \protect \BOthers {.}}{%
{\protect \APACyear {2016}}%
}]{%
bie2016}
\APACinsertmetastar {%
bie2016}%
\begin{APACrefauthors}%
Bie, R.%
, Mehmood, R.%
, Ruan, S.%
, Sun, Y.%
\BCBL {}\ \BBA {} Dawood, H.%
\end{APACrefauthors}%
\unskip\
\newblock
\APACrefYearMonthDay{2016}{}{}.
\newblock
{\BBOQ}\APACrefatitle {Adaptive fuzzy clustering by fast search and find of
  density peaks} {Adaptive fuzzy clustering by fast search and find of density
  peaks}.{\BBCQ}
\newblock
\APACjournalVolNumPages{Personal and Ubiquitous Computing}{20}{5}{785--793}.
\newblock
\begin{APACrefDOI} \doi{10.1007/s00779-016-0954-4} \end{APACrefDOI}
\PrintBackRefs{\CurrentBib}

\bibitem [\protect \citeauthoryear {%
Birant%
\ \BBA {} Kut%
}{%
Birant%
\ \BBA {} Kut%
}{%
{\protect \APACyear {2007}}%
}]{%
birant2007}
\APACinsertmetastar {%
birant2007}%
\begin{APACrefauthors}%
Birant, D.%
\BCBT {}\ \BBA {} Kut, A.%
\end{APACrefauthors}%
\unskip\
\newblock
\APACrefYearMonthDay{2007}{}{}.
\newblock
{\BBOQ}\APACrefatitle {ST-DBSCAN: An algorithm for clustering spatial--temporal
  data} {St-dbscan: An algorithm for clustering spatial--temporal data}.{\BBCQ}
\newblock
\APACjournalVolNumPages{Data \& Knowledge Engineering}{60}{1}{208--221}.
\newblock
\begin{APACrefDOI} \doi{10.1016/j.datak.2006.01.013} \end{APACrefDOI}
\PrintBackRefs{\CurrentBib}

\bibitem [\protect \citeauthoryear {%
Blangiardo%
, Finazzi%
\BCBL {}\ \BBA {} Cameletti%
}{%
Blangiardo%
\ \protect \BOthers {.}}{%
{\protect \APACyear {2016}}%
}]{%
blangiardo2016}
\APACinsertmetastar {%
blangiardo2016}%
\begin{APACrefauthors}%
Blangiardo, M.%
, Finazzi, F.%
\BCBL {}\ \BBA {} Cameletti, M.%
\end{APACrefauthors}%
\unskip\
\newblock
\APACrefYearMonthDay{2016}{}{}.
\newblock
{\BBOQ}\APACrefatitle {Two-stage Bayesian model to evaluate the effect of air
  pollution on chronic respiratory diseases using drug prescriptions}
  {Two-stage bayesian model to evaluate the effect of air pollution on chronic
  respiratory diseases using drug prescriptions}.{\BBCQ}
\newblock
\APACjournalVolNumPages{Spatial and spatio-temporal epidemiology}{18}{}{1--12}.
\newblock
\begin{APACrefDOI} \doi{10.1016/j.sste.2016.03.001} \end{APACrefDOI}
\PrintBackRefs{\CurrentBib}

\bibitem [\protect \citeauthoryear {%
Borah%
\ \BBA {} Bhattacharyya%
}{%
Borah%
\ \BBA {} Bhattacharyya%
}{%
{\protect \APACyear {2004}}%
}]{%
borah2004}
\APACinsertmetastar {%
borah2004}%
\begin{APACrefauthors}%
Borah, B.%
\BCBT {}\ \BBA {} Bhattacharyya, D.%
\end{APACrefauthors}%
\unskip\
\newblock
\APACrefYearMonthDay{2004}{}{}.
\newblock
{\BBOQ}\APACrefatitle {An improved sampling-based DBSCAN for large spatial
  databases} {An improved sampling-based dbscan for large spatial
  databases}.{\BBCQ}
\newblock
\BIn{} \APACrefbtitle {International Conference on Intelligent Sensing and
  Information Processing, 2004. Proceedings of} {International conference on
  intelligent sensing and information processing, 2004. proceedings of}\
  (\BPGS\ 92--96).
\newblock
\begin{APACrefDOI} \doi{10.1109/ICISIP.2004.1287631} \end{APACrefDOI}
\PrintBackRefs{\CurrentBib}

\bibitem [\protect \citeauthoryear {%
Borgwardt%
, Brieden%
\BCBL {}\ \BBA {} Gritzmann%
}{%
Borgwardt%
\ \protect \BOthers {.}}{%
{\protect \APACyear {2017}}%
}]{%
borgwardt2017}
\APACinsertmetastar {%
borgwardt2017}%
\begin{APACrefauthors}%
Borgwardt, S.%
, Brieden, A.%
\BCBL {}\ \BBA {} Gritzmann, P.%
\end{APACrefauthors}%
\unskip\
\newblock
\APACrefYearMonthDay{2017}{}{}.
\newblock
{\BBOQ}\APACrefatitle {An LP-based k-means algorithm for balancing weighted
  point sets} {An lp-based k-means algorithm for balancing weighted point
  sets}.{\BBCQ}
\newblock
\APACjournalVolNumPages{European Journal of Operational
  Research}{263}{2}{349--355}.
\newblock
\begin{APACrefDOI} \doi{10.1016/j.ejor.2017.04.054} \end{APACrefDOI}
\PrintBackRefs{\CurrentBib}

\bibitem [\protect \citeauthoryear {%
Cairns%
, Marshall%
\BCBL {}\ \BBA {} Kee%
}{%
Cairns%
\ \protect \BOthers {.}}{%
{\protect \APACyear {2011}}%
}]{%
cairns2011}
\APACinsertmetastar {%
cairns2011}%
\begin{APACrefauthors}%
Cairns, K\BPBI J.%
, Marshall, A\BPBI H.%
\BCBL {}\ \BBA {} Kee, F.%
\end{APACrefauthors}%
\unskip\
\newblock
\APACrefYearMonthDay{2011}{}{}.
\newblock
{\BBOQ}\APACrefatitle {Using simulation to assess cardiac first-responder
  schemes exhibiting stochastic and spatial complexities} {Using simulation to
  assess cardiac first-responder schemes exhibiting stochastic and spatial
  complexities}.{\BBCQ}
\newblock
\APACjournalVolNumPages{Journal of the Operational Research
  Society}{62}{6}{982--991}.
\PrintBackRefs{\CurrentBib}

\bibitem [\protect \citeauthoryear {%
Catlett%
, Cesario%
, Talia%
\BCBL {}\ \BBA {} Vinci%
}{%
Catlett%
\ \protect \BOthers {.}}{%
{\protect \APACyear {2019}}%
}]{%
catlett2019}
\APACinsertmetastar {%
catlett2019}%
\begin{APACrefauthors}%
Catlett, C.%
, Cesario, E.%
, Talia, D.%
\BCBL {}\ \BBA {} Vinci, A.%
\end{APACrefauthors}%
\unskip\
\newblock
\APACrefYearMonthDay{2019}{}{}.
\newblock
{\BBOQ}\APACrefatitle {Spatio-temporal crime predictions in smart cities: A
  data-driven approach and experiments} {Spatio-temporal crime predictions in
  smart cities: A data-driven approach and experiments}.{\BBCQ}
\newblock
\APACjournalVolNumPages{Pervasive and Mobile Computing}{}{}{}.
\newblock
\begin{APACrefDOI} \doi{10.1016/j.pmcj.2019.01.003} \end{APACrefDOI}
\PrintBackRefs{\CurrentBib}

\bibitem [\protect \citeauthoryear {%
Chen%
, {\"O}zsu%
\BCBL {}\ \BBA {} Oria%
}{%
Chen%
\ \protect \BOthers {.}}{%
{\protect \APACyear {2005}}%
}]{%
chen2005}
\APACinsertmetastar {%
chen2005}%
\begin{APACrefauthors}%
Chen, L.%
, {\"O}zsu, M\BPBI T.%
\BCBL {}\ \BBA {} Oria, V.%
\end{APACrefauthors}%
\unskip\
\newblock
\APACrefYearMonthDay{2005}{}{}.
\newblock
{\BBOQ}\APACrefatitle {Robust and fast similarity search for moving object
  trajectories} {Robust and fast similarity search for moving object
  trajectories}.{\BBCQ}
\newblock
\BIn{} \APACrefbtitle {Proceedings of the 2005 ACM SIGMOD international
  conference on Management of data} {Proceedings of the 2005 acm sigmod
  international conference on management of data}\ (\BPGS\ 491--502).
\newblock
\begin{APACrefDOI} \doi{10.1145/1066157.1066213} \end{APACrefDOI}
\PrintBackRefs{\CurrentBib}

\bibitem [\protect \citeauthoryear {%
Chimwayi%
\ \BBA {} Anuradha%
}{%
Chimwayi%
\ \BBA {} Anuradha%
}{%
{\protect \APACyear {2018}}%
}]{%
chimwayi2018}
\APACinsertmetastar {%
chimwayi2018}%
\begin{APACrefauthors}%
Chimwayi, K.%
\BCBT {}\ \BBA {} Anuradha, J.%
\end{APACrefauthors}%
\unskip\
\newblock
\APACrefYearMonthDay{2018}{}{}.
\newblock
{\BBOQ}\APACrefatitle {Clustering West Nile Virus Spatio-temporal data using
  ST-DBSCAN} {Clustering west nile virus spatio-temporal data using
  st-dbscan}.{\BBCQ}
\newblock
\APACjournalVolNumPages{Procedia computer science}{132}{}{1218--1227}.
\newblock
\begin{APACrefDOI} \doi{10.1016/j.procs.2018.05.037} \end{APACrefDOI}
\PrintBackRefs{\CurrentBib}

\bibitem [\protect \citeauthoryear {%
Coll%
, Moutari%
\BCBL {}\ \BBA {} Marshall%
}{%
Coll%
\ \protect \BOthers {.}}{%
{\protect \APACyear {2014}}%
}]{%
coll2014}
\APACinsertmetastar {%
coll2014}%
\begin{APACrefauthors}%
Coll, B.%
, Moutari, S.%
\BCBL {}\ \BBA {} Marshall, A.%
\end{APACrefauthors}%
\unskip\
\newblock
\APACrefYearMonthDay{2014}{}{}.
\newblock
{\BBOQ}\APACrefatitle {Pattern Recognition Approach for Road Collision Hotspot
  Analysis: Case Study of Northern Ireland'} {Pattern recognition approach for
  road collision hotspot analysis: Case study of northern ireland'}.{\BBCQ}
\newblock
\APACjournalVolNumPages{ITRN2014, University of Limerick}{}{}{}.
\PrintBackRefs{\CurrentBib}

\bibitem [\protect \citeauthoryear {%
Cressie%
}{%
Cressie%
}{%
{\protect \APACyear {1992}}%
}]{%
cressie1992}
\APACinsertmetastar {%
cressie1992}%
\begin{APACrefauthors}%
Cressie, N.%
\end{APACrefauthors}%
\unskip\
\newblock
\APACrefYearMonthDay{1992}{}{}.
\newblock
{\BBOQ}\APACrefatitle {Statistics for spatial data} {Statistics for spatial
  data}.{\BBCQ}
\newblock
\APACjournalVolNumPages{Terra Nova}{4}{5}{613--617}.
\PrintBackRefs{\CurrentBib}

\bibitem [\protect \citeauthoryear {%
De~Angelis%
\ \BBA {} Dias%
}{%
De~Angelis%
\ \BBA {} Dias%
}{%
{\protect \APACyear {2014}}%
}]{%
deAngelis2014}
\APACinsertmetastar {%
deAngelis2014}%
\begin{APACrefauthors}%
De~Angelis, L.%
\BCBT {}\ \BBA {} Dias, J\BPBI G.%
\end{APACrefauthors}%
\unskip\
\newblock
\APACrefYearMonthDay{2014}{}{}.
\newblock
{\BBOQ}\APACrefatitle {Mining categorical sequences from data using a hybrid
  clustering method} {Mining categorical sequences from data using a hybrid
  clustering method}.{\BBCQ}
\newblock
\APACjournalVolNumPages{European Journal of Operational
  Research}{234}{3}{720--730}.
\newblock
\begin{APACrefDOI} \doi{10.1016/j.ejor.2013.11.002} \end{APACrefDOI}
\PrintBackRefs{\CurrentBib}

\bibitem [\protect \citeauthoryear {%
{Department of Health and Social Care}%
}{%
{Department of Health and Social Care}%
}{%
{\protect \APACyear {2016}}%
}]{%
department2016}
\APACinsertmetastar {%
department2016}%
\begin{APACrefauthors}%
{Department of Health and Social Care}.%
\end{APACrefauthors}%
\unskip\
\newblock
\APACrefYearMonthDay{2016}{}{}.
\newblock
\APACrefbtitle {Government response to the Review on Antimicrobial Resistance}
  {Government response to the review on antimicrobial resistance}\
  \APACbVolEdTR{}{\BTR{}}.
\newblock
\begin{APACrefURL}
  \url{https://assets.publishing.service.gov.uk/government/uploads/system/uploads/attachment_data/file/553471/Gov_response_AMR_Review.pdf}
  \end{APACrefURL}
\PrintBackRefs{\CurrentBib}

\bibitem [\protect \citeauthoryear {%
Dias%
, Vermunt%
\BCBL {}\ \BBA {} Ramos%
}{%
Dias%
\ \protect \BOthers {.}}{%
{\protect \APACyear {2015}}%
}]{%
dias2015}
\APACinsertmetastar {%
dias2015}%
\begin{APACrefauthors}%
Dias, J\BPBI G.%
, Vermunt, J\BPBI K.%
\BCBL {}\ \BBA {} Ramos, S.%
\end{APACrefauthors}%
\unskip\
\newblock
\APACrefYearMonthDay{2015}{}{}.
\newblock
{\BBOQ}\APACrefatitle {Clustering financial time series: New insights from an
  extended hidden Markov model} {Clustering financial time series: New insights
  from an extended hidden markov model}.{\BBCQ}
\newblock
\APACjournalVolNumPages{European Journal of Operational
  Research}{243}{3}{852--864}.
\newblock
\begin{APACrefDOI} \doi{10.1016/j.ejor.2014.12.041} \end{APACrefDOI}
\PrintBackRefs{\CurrentBib}

\bibitem [\protect \citeauthoryear {%
Di~Lascio%
, Durante%
\BCBL {}\ \BBA {} Pappada%
}{%
Di~Lascio%
\ \protect \BOthers {.}}{%
{\protect \APACyear {2017}}%
}]{%
diLascio2017}
\APACinsertmetastar {%
diLascio2017}%
\begin{APACrefauthors}%
Di~Lascio, F\BPBI M\BPBI L.%
, Durante, F.%
\BCBL {}\ \BBA {} Pappada, R.%
\end{APACrefauthors}%
\unskip\
\newblock
\APACrefYearMonthDay{2017}{}{}.
\newblock
{\BBOQ}\APACrefatitle {Copula--based clustering methods} {Copula--based
  clustering methods}.{\BBCQ}
\newblock
\BIn{} \APACrefbtitle {Copulas and Dependence Models with Applications}
  {Copulas and dependence models with applications}\ (\BPGS\ 49--67).
\newblock
\APACaddressPublisher{}{Springer}.
\newblock
\begin{APACrefDOI} \doi{10.1007/978-3-319-64221-5_4} \end{APACrefDOI}
\PrintBackRefs{\CurrentBib}

\bibitem [\protect \citeauthoryear {%
Disegna%
, D’Urso%
\BCBL {}\ \BBA {} Durante%
}{%
Disegna%
\ \protect \BOthers {.}}{%
{\protect \APACyear {2017}}%
}]{%
disegna2017}
\APACinsertmetastar {%
disegna2017}%
\begin{APACrefauthors}%
Disegna, M.%
, D’Urso, P.%
\BCBL {}\ \BBA {} Durante, F.%
\end{APACrefauthors}%
\unskip\
\newblock
\APACrefYearMonthDay{2017}{}{}.
\newblock
{\BBOQ}\APACrefatitle {Copula-based fuzzy clustering of spatial time series}
  {Copula-based fuzzy clustering of spatial time series}.{\BBCQ}
\newblock
\APACjournalVolNumPages{Spatial Statistics}{21}{}{209--225}.
\newblock
\begin{APACrefDOI} \doi{10.1016/j.spasta.2017.07.002} \end{APACrefDOI}
\PrintBackRefs{\CurrentBib}

\bibitem [\protect \citeauthoryear {%
Dougherty%
\ \protect \BOthers {.}}{%
Dougherty%
\ \protect \BOthers {.}}{%
{\protect \APACyear {2002}}%
}]{%
dougherty2002}
\APACinsertmetastar {%
dougherty2002}%
\begin{APACrefauthors}%
Dougherty, E\BPBI R.%
, Barrera, J.%
, Brun, M.%
, Kim, S.%
, Cesar, R\BPBI M.%
, Chen, Y.%
\BDBL {}Trent, J\BPBI M.%
\end{APACrefauthors}%
\unskip\
\newblock
\APACrefYearMonthDay{2002}{}{}.
\newblock
{\BBOQ}\APACrefatitle {Inference from clustering with application to
  gene-expression microarrays} {Inference from clustering with application to
  gene-expression microarrays}.{\BBCQ}
\newblock
\APACjournalVolNumPages{Journal of Computational Biology}{9}{1}{105--126}.
\newblock
\begin{APACrefDOI} \doi{10.1089/10665270252833217} \end{APACrefDOI}
\PrintBackRefs{\CurrentBib}

\bibitem [\protect \citeauthoryear {%
Dupor%
\ \BBA {} McCrory%
}{%
Dupor%
\ \BBA {} McCrory%
}{%
{\protect \APACyear {2017}}%
}]{%
dupor2017}
\APACinsertmetastar {%
dupor2017}%
\begin{APACrefauthors}%
Dupor, B.%
\BCBT {}\ \BBA {} McCrory, P\BPBI B.%
\end{APACrefauthors}%
\unskip\
\newblock
\APACrefYearMonthDay{2017}{}{}.
\newblock
{\BBOQ}\APACrefatitle {A Cup Runneth Over: Fiscal Policy Spillovers from the
  2009 Recovery Act} {A cup runneth over: Fiscal policy spillovers from the
  2009 recovery act}.{\BBCQ}
\newblock
\APACjournalVolNumPages{The Economic Journal}{128}{611}{1476--1508}.
\newblock
\begin{APACrefDOI} \doi{10.1111/ecoj.12475} \end{APACrefDOI}
\PrintBackRefs{\CurrentBib}

\bibitem [\protect \citeauthoryear {%
Durkin%
\ \protect \BOthers {.}}{%
Durkin%
\ \protect \BOthers {.}}{%
{\protect \APACyear {2018}}%
}]{%
durkin2018}
\APACinsertmetastar {%
durkin2018}%
\begin{APACrefauthors}%
Durkin, M\BPBI J.%
, Jafarzadeh, S\BPBI R.%
, Hsueh, K.%
, Sallah, Y\BPBI H.%
, Munshi, K\BPBI D.%
, Henderson, R\BPBI R.%
\BCBL {}\ \BBA {} Fraser, V\BPBI J.%
\end{APACrefauthors}%
\unskip\
\newblock
\APACrefYearMonthDay{2018}{}{}.
\newblock
{\BBOQ}\APACrefatitle {Outpatient antibiotic prescription trends in the United
  States: a national cohort study} {Outpatient antibiotic prescription trends
  in the united states: a national cohort study}.{\BBCQ}
\newblock
\APACjournalVolNumPages{Infection Control \& Hospital
  Epidemiology}{39}{5}{584--589}.
\PrintBackRefs{\CurrentBib}

\bibitem [\protect \citeauthoryear {%
Ert{\"o}z%
, Steinbach%
\BCBL {}\ \BBA {} Kumar%
}{%
Ert{\"o}z%
\ \protect \BOthers {.}}{%
{\protect \APACyear {2003}}%
}]{%
ertoez2003}
\APACinsertmetastar {%
ertoez2003}%
\begin{APACrefauthors}%
Ert{\"o}z, L.%
, Steinbach, M.%
\BCBL {}\ \BBA {} Kumar, V.%
\end{APACrefauthors}%
\unskip\
\newblock
\APACrefYearMonthDay{2003}{}{}.
\newblock
{\BBOQ}\APACrefatitle {Finding clusters of different sizes, shapes, and
  densities in noisy, high dimensional data} {Finding clusters of different
  sizes, shapes, and densities in noisy, high dimensional data}.{\BBCQ}
\newblock
\BIn{} \APACrefbtitle {Proceedings of the 2003 SIAM international conference on
  data mining} {Proceedings of the 2003 siam international conference on data
  mining}\ (\BPGS\ 47--58).
\newblock
\begin{APACrefDOI} \doi{10.1137/1.9781611972733.5} \end{APACrefDOI}
\PrintBackRefs{\CurrentBib}

\bibitem [\protect \citeauthoryear {%
Eslinger%
\ \BBA {} Morgan%
}{%
Eslinger%
\ \BBA {} Morgan%
}{%
{\protect \APACyear {2017}}%
}]{%
eslinger2017}
\APACinsertmetastar {%
eslinger2017}%
\begin{APACrefauthors}%
Eslinger, R.%
\BCBT {}\ \BBA {} Morgan, J\BPBI D.%
\end{APACrefauthors}%
\unskip\
\newblock
\APACrefYearMonthDay{2017}{}{}.
\newblock
{\BBOQ}\APACrefatitle {Spatial Cluster Analysis of High-Density Vehicle--Bear
  Collisions and Bridge Locations} {Spatial cluster analysis of high-density
  vehicle--bear collisions and bridge locations}.{\BBCQ}
\newblock
\APACjournalVolNumPages{Papers in Applied Geography}{3}{2}{171--181}.
\newblock
\begin{APACrefDOI} \doi{10.1080/23754931.2017.1299633} \end{APACrefDOI}
\PrintBackRefs{\CurrentBib}

\bibitem [\protect \citeauthoryear {%
Ester%
, Kriegel%
, Sander%
, Xu%
\BCBL {}\ \protect \BOthers {.}}{%
Ester%
\ \protect \BOthers {.}}{%
{\protect \APACyear {1996}}%
}]{%
ester1996}
\APACinsertmetastar {%
ester1996}%
\begin{APACrefauthors}%
Ester, M.%
, Kriegel, H\BHBI P.%
, Sander, J.%
, Xu, X.%
\BCBL {}\ \BOthersPeriod {.}\end{APACrefauthors}%
\unskip\
\newblock
\APACrefYearMonthDay{1996}{}{}.
\newblock
{\BBOQ}\APACrefatitle {A density-based algorithm for discovering clusters in
  large spatial databases with noise.} {A density-based algorithm for
  discovering clusters in large spatial databases with noise.}{\BBCQ}
\newblock
\BIn{} \APACrefbtitle {Kdd} {Kdd}\ (\BVOL~96, \BPGS\ 226--231).
\PrintBackRefs{\CurrentBib}

\bibitem [\protect \citeauthoryear {%
Everitt%
}{%
Everitt%
}{%
{\protect \APACyear {2011}}%
}]{%
everitt2011}
\APACinsertmetastar {%
everitt2011}%
\begin{APACrefauthors}%
Everitt, B.%
\end{APACrefauthors}%
\unskip\
\newblock
\APACrefYear{2011}.
\newblock
\APACrefbtitle {Cluster Analysis} {Cluster analysis}\ (\PrintOrdinal{5th
  edition.}\ \BEd).
\newblock
\APACaddressPublisher{Hoboken}{Wiley}.
\PrintBackRefs{\CurrentBib}

\bibitem [\protect \citeauthoryear {%
Gerber%
}{%
Gerber%
}{%
{\protect \APACyear {2014}}%
}]{%
gerber2014}
\APACinsertmetastar {%
gerber2014}%
\begin{APACrefauthors}%
Gerber, M\BPBI S.%
\end{APACrefauthors}%
\unskip\
\newblock
\APACrefYearMonthDay{2014}{}{}.
\newblock
{\BBOQ}\APACrefatitle {Predicting crime using Twitter and kernel density
  estimation} {Predicting crime using twitter and kernel density
  estimation}.{\BBCQ}
\newblock
\APACjournalVolNumPages{Decision Support Systems}{61}{}{115--125}.
\newblock
\begin{APACrefDOI} \doi{10.1016/j.dss.2014.02.003} \end{APACrefDOI}
\PrintBackRefs{\CurrentBib}

\bibitem [\protect \citeauthoryear {%
Gomide%
\ \protect \BOthers {.}}{%
Gomide%
\ \protect \BOthers {.}}{%
{\protect \APACyear {2011}}%
}]{%
gomide2011}
\APACinsertmetastar {%
gomide2011}%
\begin{APACrefauthors}%
Gomide, J.%
, Veloso, A.%
, Meira~Jr, W.%
, Almeida, V.%
, Benevenuto, F.%
, Ferraz, F.%
\BCBL {}\ \BBA {} Teixeira, M.%
\end{APACrefauthors}%
\unskip\
\newblock
\APACrefYearMonthDay{2011}{}{}.
\newblock
{\BBOQ}\APACrefatitle {Dengue surveillance based on a computational model of
  spatio-temporal locality of Twitter} {Dengue surveillance based on a
  computational model of spatio-temporal locality of twitter}.{\BBCQ}
\newblock
\BIn{} \APACrefbtitle {Proceedings of the 3rd international web science
  conference} {Proceedings of the 3rd international web science conference}\
  (\BPG~3).
\newblock
\begin{APACrefDOI} \doi{10.1145/2527031.2527049} \end{APACrefDOI}
\PrintBackRefs{\CurrentBib}

\bibitem [\protect \citeauthoryear {%
Han%
, Kamber%
\BCBL {}\ \BBA {} Pei%
}{%
Han%
\ \protect \BOthers {.}}{%
{\protect \APACyear {2011}}%
}]{%
han2011}
\APACinsertmetastar {%
han2011}%
\begin{APACrefauthors}%
Han, J.%
, Kamber, M.%
\BCBL {}\ \BBA {} Pei, J.%
\end{APACrefauthors}%
\unskip\
\newblock
\APACrefYear{2011}.
\newblock
\APACrefbtitle {Data Mining: Concepts and Techniques} {Data mining: Concepts
  and techniques}\ (\PrintOrdinal{3}\ \BEd).
\newblock
\APACaddressPublisher{}{Elsevier Science}.
\PrintBackRefs{\CurrentBib}

\bibitem [\protect \citeauthoryear {%
Hassin%
\ \BBA {} Or%
}{%
Hassin%
\ \BBA {} Or%
}{%
{\protect \APACyear {2010}}%
}]{%
hassin2010}
\APACinsertmetastar {%
hassin2010}%
\begin{APACrefauthors}%
Hassin, R.%
\BCBT {}\ \BBA {} Or, E.%
\end{APACrefauthors}%
\unskip\
\newblock
\APACrefYearMonthDay{2010}{}{}.
\newblock
{\BBOQ}\APACrefatitle {Min sum clustering with penalties} {Min sum clustering
  with penalties}.{\BBCQ}
\newblock
\APACjournalVolNumPages{European Journal of Operational
  Research}{206}{3}{547--554}.
\newblock
\begin{APACrefDOI} \doi{10.1016/j.ejor.2010.03.004} \end{APACrefDOI}
\PrintBackRefs{\CurrentBib}

\bibitem [\protect \citeauthoryear {%
{Healthcare Quality and Improvement Directorate}%
}{%
{Healthcare Quality and Improvement Directorate}%
}{%
{\protect \APACyear {2018}}%
}]{%
healthcare2018}
\APACinsertmetastar {%
healthcare2018}%
\begin{APACrefauthors}%
{Healthcare Quality and Improvement Directorate}.%
\end{APACrefauthors}%
\unskip\
\newblock
\APACrefYearMonthDay{2018}{}{}.
\newblock
\APACrefbtitle {Practising Realistic Medicine: Chief Medical Officer for
  Scotland annual report} {Practising realistic medicine: Chief medical officer
  for scotland annual report}\ \APACbVolEdTR{}{\BTR{}}.
\newblock
\begin{APACrefURL}
  \url{https://www.gov.scot/publications/practising-realistic-medicine/}
  \end{APACrefURL}
\PrintBackRefs{\CurrentBib}

\bibitem [\protect \citeauthoryear {%
Hijmans%
}{%
Hijmans%
}{%
{\protect \APACyear {2017}}%
}]{%
hijmans2017}
\APACinsertmetastar {%
hijmans2017}%
\begin{APACrefauthors}%
Hijmans, R\BPBI J.%
\end{APACrefauthors}%
\unskip\
\newblock
\APACrefYearMonthDay{2017}{}{}.
\newblock
{\BBOQ}\APACrefatitle {geosphere: Spherical Trigonometry} {geosphere: Spherical
  trigonometry}{\BBCQ}\ [\bibcomputersoftwaremanual].
\newblock
\begin{APACrefURL} \url{https://CRAN.R-project.org/package=geosphere}
  \end{APACrefURL}
\newblock
\APACrefnote{R package version 1.5-7}
\PrintBackRefs{\CurrentBib}

\bibitem [\protect \citeauthoryear {%
Hinneburg%
\ \BBA {} Gabriel%
}{%
Hinneburg%
\ \BBA {} Gabriel%
}{%
{\protect \APACyear {2007}}%
}]{%
hinneburg2007}
\APACinsertmetastar {%
hinneburg2007}%
\begin{APACrefauthors}%
Hinneburg, A.%
\BCBT {}\ \BBA {} Gabriel, H\BHBI H.%
\end{APACrefauthors}%
\unskip\
\newblock
\APACrefYearMonthDay{2007}{}{}.
\newblock
{\BBOQ}\APACrefatitle {Denclue 2.0: Fast clustering based on kernel density
  estimation} {Denclue 2.0: Fast clustering based on kernel density
  estimation}.{\BBCQ}
\newblock
\BIn{} \APACrefbtitle {International symposium on intelligent data analysis}
  {International symposium on intelligent data analysis}\ (\BPGS\ 70--80).
\newblock
\begin{APACrefDOI} \doi{10.1007/978-3-540-74825-0_7} \end{APACrefDOI}
\PrintBackRefs{\CurrentBib}

\bibitem [\protect \citeauthoryear {%
Inniss%
}{%
Inniss%
}{%
{\protect \APACyear {2006}}%
}]{%
inniss2006}
\APACinsertmetastar {%
inniss2006}%
\begin{APACrefauthors}%
Inniss, T\BPBI R.%
\end{APACrefauthors}%
\unskip\
\newblock
\APACrefYearMonthDay{2006}{}{}.
\newblock
{\BBOQ}\APACrefatitle {Seasonal clustering technique for time series data}
  {Seasonal clustering technique for time series data}.{\BBCQ}
\newblock
\APACjournalVolNumPages{European Journal of Operational
  Research}{175}{1}{376--384}.
\newblock
\begin{APACrefDOI} \doi{10.1016/j.ejor.2005.03.049} \end{APACrefDOI}
\PrintBackRefs{\CurrentBib}

\bibitem [\protect \citeauthoryear {%
Izakian%
, Pedrycz%
\BCBL {}\ \BBA {} Jamal%
}{%
Izakian%
\ \protect \BOthers {.}}{%
{\protect \APACyear {2015}}%
}]{%
izakian2015}
\APACinsertmetastar {%
izakian2015}%
\begin{APACrefauthors}%
Izakian, H.%
, Pedrycz, W.%
\BCBL {}\ \BBA {} Jamal, I.%
\end{APACrefauthors}%
\unskip\
\newblock
\APACrefYearMonthDay{2015}{}{}.
\newblock
{\BBOQ}\APACrefatitle {Fuzzy clustering of time series data using dynamic time
  warping distance} {Fuzzy clustering of time series data using dynamic time
  warping distance}.{\BBCQ}
\newblock
\APACjournalVolNumPages{Engineering Applications of Artificial
  Intelligence}{39}{}{235--244}.
\newblock
\begin{APACrefDOI} \doi{10.1016/j.engappai.2014.12.015} \end{APACrefDOI}
\PrintBackRefs{\CurrentBib}

\bibitem [\protect \citeauthoryear {%
Katz%
\ \protect \BOthers {.}}{%
Katz%
\ \protect \BOthers {.}}{%
{\protect \APACyear {2010}}%
}]{%
katz2010}
\APACinsertmetastar {%
katz2010}%
\begin{APACrefauthors}%
Katz, N.%
, Panas, L.%
, Kim, M.%
, Audet, A\BPBI D.%
, Bilansky, A.%
, Eadie, J.%
\BDBL {}Carrow, G.%
\end{APACrefauthors}%
\unskip\
\newblock
\APACrefYearMonthDay{2010}{}{}.
\newblock
{\BBOQ}\APACrefatitle {Usefulness of prescription monitoring programs for
  surveillance—analysis of Schedule II opioid prescription data in
  Massachusetts, 1996--2006} {Usefulness of prescription monitoring programs
  for surveillance—analysis of schedule ii opioid prescription data in
  massachusetts, 1996--2006}.{\BBCQ}
\newblock
\APACjournalVolNumPages{Pharmacoepidemiology and drug safety}{19}{2}{115--123}.
\newblock
\begin{APACrefDOI} \doi{10.1002/pds.1878} \end{APACrefDOI}
\PrintBackRefs{\CurrentBib}

\bibitem [\protect \citeauthoryear {%
Kaufman%
\ \BBA {} Rousseeuw%
}{%
Kaufman%
\ \BBA {} Rousseeuw%
}{%
{\protect \APACyear {2009}}%
}]{%
kaufman2009}
\APACinsertmetastar {%
kaufman2009}%
\begin{APACrefauthors}%
Kaufman, L.%
\BCBT {}\ \BBA {} Rousseeuw, P\BPBI J.%
\end{APACrefauthors}%
\unskip\
\newblock
\APACrefYear{2009}.
\newblock
\APACrefbtitle {Finding groups in data: an introduction to cluster analysis}
  {Finding groups in data: an introduction to cluster analysis}\ (\BVOL~3).
\newblock
\APACaddressPublisher{}{John Wiley \& Sons}.
\PrintBackRefs{\CurrentBib}

\bibitem [\protect \citeauthoryear {%
Kiselev%
, Andrews%
\BCBL {}\ \BBA {} Hemberg%
}{%
Kiselev%
\ \protect \BOthers {.}}{%
{\protect \APACyear {2019}}%
}]{%
kiselev2019}
\APACinsertmetastar {%
kiselev2019}%
\begin{APACrefauthors}%
Kiselev, V\BPBI Y.%
, Andrews, T\BPBI S.%
\BCBL {}\ \BBA {} Hemberg, M.%
\end{APACrefauthors}%
\unskip\
\newblock
\APACrefYearMonthDay{2019}{}{}.
\newblock
{\BBOQ}\APACrefatitle {Challenges in unsupervised clustering of single-cell
  RNA-seq data} {Challenges in unsupervised clustering of single-cell rna-seq
  data}.{\BBCQ}
\newblock
\APACjournalVolNumPages{Nature Reviews Genetics}{}{}{1}.
\newblock
\begin{APACrefDOI} \doi{10.1038/s41576-018-0088-9} \end{APACrefDOI}
\PrintBackRefs{\CurrentBib}

\bibitem [\protect \citeauthoryear {%
Kj{\ae}rulff%
, Ersb{\o}ll%
, Gislason%
\BCBL {}\ \BBA {} Schipperijn%
}{%
Kj{\ae}rulff%
\ \protect \BOthers {.}}{%
{\protect \APACyear {2016}}%
}]{%
kjaerulff2016}
\APACinsertmetastar {%
kjaerulff2016}%
\begin{APACrefauthors}%
Kj{\ae}rulff, T\BPBI M.%
, Ersb{\o}ll, A\BPBI K.%
, Gislason, G.%
\BCBL {}\ \BBA {} Schipperijn, J.%
\end{APACrefauthors}%
\unskip\
\newblock
\APACrefYearMonthDay{2016}{}{}.
\newblock
{\BBOQ}\APACrefatitle {Geographical clustering of incident acute myocardial
  infarction in Denmark: A spatial analysis approach} {Geographical clustering
  of incident acute myocardial infarction in denmark: A spatial analysis
  approach}.{\BBCQ}
\newblock
\APACjournalVolNumPages{Spatial and spatio-temporal
  epidemiology}{19}{}{46--59}.
\newblock
\begin{APACrefDOI} \doi{10.1016/j.sste.2016.05.001} \end{APACrefDOI}
\PrintBackRefs{\CurrentBib}

\bibitem [\protect \citeauthoryear {%
Kriegel%
, Kr{\"o}ger%
, Sander%
\BCBL {}\ \BBA {} Zimek%
}{%
Kriegel%
\ \protect \BOthers {.}}{%
{\protect \APACyear {2011}}%
}]{%
kriegel2011}
\APACinsertmetastar {%
kriegel2011}%
\begin{APACrefauthors}%
Kriegel, H\BHBI P.%
, Kr{\"o}ger, P.%
, Sander, J.%
\BCBL {}\ \BBA {} Zimek, A.%
\end{APACrefauthors}%
\unskip\
\newblock
\APACrefYearMonthDay{2011}{}{}.
\newblock
{\BBOQ}\APACrefatitle {Density-based clustering} {Density-based
  clustering}.{\BBCQ}
\newblock
\APACjournalVolNumPages{Wiley Interdisciplinary Reviews: Data Mining and
  Knowledge Discovery}{1}{3}{231--240}.
\newblock
\begin{APACrefDOI} \doi{10.1002/widm.30} \end{APACrefDOI}
\PrintBackRefs{\CurrentBib}

\bibitem [\protect \citeauthoryear {%
Liao%
}{%
Liao%
}{%
{\protect \APACyear {2005}}%
}]{%
liao2005}
\APACinsertmetastar {%
liao2005}%
\begin{APACrefauthors}%
Liao, T\BPBI W.%
\end{APACrefauthors}%
\unskip\
\newblock
\APACrefYearMonthDay{2005}{}{}.
\newblock
{\BBOQ}\APACrefatitle {Clustering of time series data—a survey} {Clustering
  of time series data—a survey}.{\BBCQ}
\newblock
\APACjournalVolNumPages{Pattern recognition}{38}{11}{1857--1874}.
\newblock
\begin{APACrefDOI} \doi{10.1016/j.patcog.2005.01.025} \end{APACrefDOI}
\PrintBackRefs{\CurrentBib}

\bibitem [\protect \citeauthoryear {%
Mai%
, Fry%
\BCBL {}\ \BBA {} Ohlmann%
}{%
Mai%
\ \protect \BOthers {.}}{%
{\protect \APACyear {2018}}%
}]{%
mai2018}
\APACinsertmetastar {%
mai2018}%
\begin{APACrefauthors}%
Mai, F.%
, Fry, M\BPBI J.%
\BCBL {}\ \BBA {} Ohlmann, J\BPBI W.%
\end{APACrefauthors}%
\unskip\
\newblock
\APACrefYearMonthDay{2018}{}{}.
\newblock
{\BBOQ}\APACrefatitle {Model-based capacitated clustering with posterior
  regularization} {Model-based capacitated clustering with posterior
  regularization}.{\BBCQ}
\newblock
\APACjournalVolNumPages{European Journal of Operational
  Research}{271}{2}{594--605}.
\newblock
\begin{APACrefDOI} \doi{10.1016/j.ejor.2018.04.048} \end{APACrefDOI}
\PrintBackRefs{\CurrentBib}

\bibitem [\protect \citeauthoryear {%
Marbac%
, Biernacki%
\BCBL {}\ \BBA {} Vandewalle%
}{%
Marbac%
\ \protect \BOthers {.}}{%
{\protect \APACyear {2017}}%
}]{%
marbac2017}
\APACinsertmetastar {%
marbac2017}%
\begin{APACrefauthors}%
Marbac, M.%
, Biernacki, C.%
\BCBL {}\ \BBA {} Vandewalle, V.%
\end{APACrefauthors}%
\unskip\
\newblock
\APACrefYearMonthDay{2017}{}{}.
\newblock
{\BBOQ}\APACrefatitle {Model-based clustering of Gaussian copulas for mixed
  data} {Model-based clustering of gaussian copulas for mixed data}.{\BBCQ}
\newblock
\APACjournalVolNumPages{Communications in Statistics-Theory and
  Methods}{46}{23}{11635--11656}.
\newblock
\begin{APACrefDOI} \doi{10.1080/03610926.2016.1277753} \end{APACrefDOI}
\PrintBackRefs{\CurrentBib}

\bibitem [\protect \citeauthoryear {%
Matioli%
, Santos%
, Kleina%
\BCBL {}\ \BBA {} Leite%
}{%
Matioli%
\ \protect \BOthers {.}}{%
{\protect \APACyear {2018}}%
}]{%
matioli2018}
\APACinsertmetastar {%
matioli2018}%
\begin{APACrefauthors}%
Matioli, L.%
, Santos, S.%
, Kleina, M.%
\BCBL {}\ \BBA {} Leite, E.%
\end{APACrefauthors}%
\unskip\
\newblock
\APACrefYearMonthDay{2018}{}{}.
\newblock
{\BBOQ}\APACrefatitle {A new algorithm for clustering based on kernel density
  estimation} {A new algorithm for clustering based on kernel density
  estimation}.{\BBCQ}
\newblock
\APACjournalVolNumPages{Journal of Applied Statistics}{45}{2}{347--366}.
\newblock
\begin{APACrefDOI} \doi{10.1080/02664763.2016.1277191} \end{APACrefDOI}
\PrintBackRefs{\CurrentBib}

\bibitem [\protect \citeauthoryear {%
Mehmood%
, Zhang%
, Bie%
, Dawood%
\BCBL {}\ \BBA {} Ahmad%
}{%
Mehmood%
\ \protect \BOthers {.}}{%
{\protect \APACyear {2016}}%
}]{%
mehmood2016}
\APACinsertmetastar {%
mehmood2016}%
\begin{APACrefauthors}%
Mehmood, R.%
, Zhang, G.%
, Bie, R.%
, Dawood, H.%
\BCBL {}\ \BBA {} Ahmad, H.%
\end{APACrefauthors}%
\unskip\
\newblock
\APACrefYearMonthDay{2016}{}{}.
\newblock
{\BBOQ}\APACrefatitle {Clustering by fast search and find of density peaks via
  heat diffusion} {Clustering by fast search and find of density peaks via heat
  diffusion}.{\BBCQ}
\newblock
\APACjournalVolNumPages{Neurocomputing}{208}{}{210--217}.
\newblock
\begin{APACrefDOI} \doi{10.1016/j.neucom.2016.01.102} \end{APACrefDOI}
\PrintBackRefs{\CurrentBib}

\bibitem [\protect \citeauthoryear {%
Menafoglio%
\ \BBA {} Secchi%
}{%
Menafoglio%
\ \BBA {} Secchi%
}{%
{\protect \APACyear {2017}}%
}]{%
menafoglio2017}
\APACinsertmetastar {%
menafoglio2017}%
\begin{APACrefauthors}%
Menafoglio, A.%
\BCBT {}\ \BBA {} Secchi, P.%
\end{APACrefauthors}%
\unskip\
\newblock
\APACrefYearMonthDay{2017}{}{}.
\newblock
{\BBOQ}\APACrefatitle {Statistical analysis of complex and spatially dependent
  data: A review of Object Oriented Spatial Statistics} {Statistical analysis
  of complex and spatially dependent data: A review of object oriented spatial
  statistics}.{\BBCQ}
\newblock
\APACjournalVolNumPages{European journal of operational
  research}{258}{2}{401--410}.
\newblock
\begin{APACrefDOI} \doi{10.1016/j.ejor.2016.09.061} \end{APACrefDOI}
\PrintBackRefs{\CurrentBib}

\bibitem [\protect \citeauthoryear {%
M{\"o}lter%
\ \protect \BOthers {.}}{%
M{\"o}lter%
\ \protect \BOthers {.}}{%
{\protect \APACyear {2018}}%
}]{%
molter2018}
\APACinsertmetastar {%
molter2018}%
\begin{APACrefauthors}%
M{\"o}lter, A.%
, Belmonte, M.%
, Palin, V.%
, Mistry, C.%
, Sperrin, M.%
, White, A.%
\BDBL {}Van~Staa, T.%
\end{APACrefauthors}%
\unskip\
\newblock
\APACrefYearMonthDay{2018}{}{}.
\newblock
{\BBOQ}\APACrefatitle {Antibiotic prescribing patterns in general medical
  practices in England: Does area matter?} {Antibiotic prescribing patterns in
  general medical practices in england: Does area matter?}{\BBCQ}
\newblock
\APACjournalVolNumPages{Health \& place}{53}{}{10--16}.
\newblock
\begin{APACrefDOI} \doi{10.1016/j.healthplace.2018.07.004} \end{APACrefDOI}
\PrintBackRefs{\CurrentBib}

\bibitem [\protect \citeauthoryear {%
Montani%
, Portinale%
, Leonardi%
, Bellazzi%
\BCBL {}\ \BBA {} Bellazzi%
}{%
Montani%
\ \protect \BOthers {.}}{%
{\protect \APACyear {2006}}%
}]{%
montani2006}
\APACinsertmetastar {%
montani2006}%
\begin{APACrefauthors}%
Montani, S.%
, Portinale, L.%
, Leonardi, G.%
, Bellazzi, R.%
\BCBL {}\ \BBA {} Bellazzi, R.%
\end{APACrefauthors}%
\unskip\
\newblock
\APACrefYearMonthDay{2006}{}{}.
\newblock
{\BBOQ}\APACrefatitle {Case-based retrieval to support the treatment of end
  stage renal failure patients} {Case-based retrieval to support the treatment
  of end stage renal failure patients}.{\BBCQ}
\newblock
\APACjournalVolNumPages{Artificial Intelligence in Medicine}{37}{1}{31--42}.
\newblock
\begin{APACrefDOI} \doi{10.1016/j.artmed.2005.06.003} \end{APACrefDOI}
\PrintBackRefs{\CurrentBib}

\bibitem [\protect \citeauthoryear {%
Mori%
, Mendiburu%
\BCBL {}\ \BBA {} Lozano%
}{%
Mori%
\ \protect \BOthers {.}}{%
{\protect \APACyear {2016}}%
}]{%
mori2016}
\APACinsertmetastar {%
mori2016}%
\begin{APACrefauthors}%
Mori, U.%
, Mendiburu, A.%
\BCBL {}\ \BBA {} Lozano, J\BPBI A.%
\end{APACrefauthors}%
\unskip\
\newblock
\APACrefYearMonthDay{2016}{}{}.
\newblock
{\BBOQ}\APACrefatitle {Distance measures for time series in R: The TSdist
  package} {Distance measures for time series in r: The tsdist package}.{\BBCQ}
\newblock
\APACjournalVolNumPages{R Journal}{8}{2}{451--459}.
\PrintBackRefs{\CurrentBib}

\bibitem [\protect \citeauthoryear {%
Nascimento%
\ \BBA {} De~Carvalho%
}{%
Nascimento%
\ \BBA {} De~Carvalho%
}{%
{\protect \APACyear {2011}}%
}]{%
nascimento2011}
\APACinsertmetastar {%
nascimento2011}%
\begin{APACrefauthors}%
Nascimento, M\BPBI C.%
\BCBT {}\ \BBA {} De~Carvalho, A\BPBI C.%
\end{APACrefauthors}%
\unskip\
\newblock
\APACrefYearMonthDay{2011}{}{}.
\newblock
{\BBOQ}\APACrefatitle {Spectral methods for graph clustering--a survey}
  {Spectral methods for graph clustering--a survey}.{\BBCQ}
\newblock
\APACjournalVolNumPages{European Journal of Operational
  Research}{211}{2}{221--231}.
\newblock
\begin{APACrefDOI} \doi{10.1016/j.ejor.2010.08.012} \end{APACrefDOI}
\PrintBackRefs{\CurrentBib}

\bibitem [\protect \citeauthoryear {%
Ng%
\ \BBA {} Han%
}{%
Ng%
\ \BBA {} Han%
}{%
{\protect \APACyear {2002}}%
}]{%
ng2002}
\APACinsertmetastar {%
ng2002}%
\begin{APACrefauthors}%
Ng, R\BPBI T.%
\BCBT {}\ \BBA {} Han, J.%
\end{APACrefauthors}%
\unskip\
\newblock
\APACrefYearMonthDay{2002}{}{}.
\newblock
{\BBOQ}\APACrefatitle {CLARANS: A method for clustering objects for spatial
  data mining} {Clarans: A method for clustering objects for spatial data
  mining}.{\BBCQ}
\newblock
\APACjournalVolNumPages{IEEE Transactions on Knowledge \& Data
  Engineering}{}{5}{1003--1016}.
\newblock
\begin{APACrefDOI} \doi{10.1109/TKDE.2002.1033770} \end{APACrefDOI}
\PrintBackRefs{\CurrentBib}

\bibitem [\protect \citeauthoryear {%
{NHS ISD}%
}{%
{NHS ISD}%
}{%
{\protect \APACyear {2018}}%
}]{%
nhsisd2018}
\APACinsertmetastar {%
nhsisd2018}%
\begin{APACrefauthors}%
{NHS ISD}.%
\end{APACrefauthors}%
\unskip\
\newblock
\APACrefYearMonthDay{2018}{}{}.
\newblock
\APACrefbtitle {General Practice - GP Workforce and practice list sizes
  2008-2018} {General practice - gp workforce and practice list sizes
  2008-2018}\ \APACbVolEdTR{}{\BTR{}}.
\newblock
\begin{APACrefURL}
  \url{http://www.isdscotland.org/Health-Topics/General-Practice/Publications/2018-12-11/2018-12-11-GPWorkforce2018-Summary.pdf}
  \end{APACrefURL}
\PrintBackRefs{\CurrentBib}

\bibitem [\protect \citeauthoryear {%
Paparrizos%
\ \BBA {} Gravano%
}{%
Paparrizos%
\ \BBA {} Gravano%
}{%
{\protect \APACyear {2017}}%
}]{%
paparrizos2017}
\APACinsertmetastar {%
paparrizos2017}%
\begin{APACrefauthors}%
Paparrizos, J.%
\BCBT {}\ \BBA {} Gravano, L.%
\end{APACrefauthors}%
\unskip\
\newblock
\APACrefYearMonthDay{2017}{}{}.
\newblock
{\BBOQ}\APACrefatitle {Fast and accurate time-series clustering} {Fast and
  accurate time-series clustering}.{\BBCQ}
\newblock
\APACjournalVolNumPages{ACM Transactions on Database Systems (TODS)}{42}{2}{8}.
\newblock
\begin{APACrefDOI} \doi{10.1145/3044711} \end{APACrefDOI}
\PrintBackRefs{\CurrentBib}

\bibitem [\protect \citeauthoryear {%
Pei%
, Jasra%
, Hand%
, Zhu%
\BCBL {}\ \BBA {} Zhou%
}{%
Pei%
\ \protect \BOthers {.}}{%
{\protect \APACyear {2009}}%
}]{%
pei2009}
\APACinsertmetastar {%
pei2009}%
\begin{APACrefauthors}%
Pei, T.%
, Jasra, A.%
, Hand, D\BPBI J.%
, Zhu, A\BHBI X.%
\BCBL {}\ \BBA {} Zhou, C.%
\end{APACrefauthors}%
\unskip\
\newblock
\APACrefYearMonthDay{2009}{}{}.
\newblock
{\BBOQ}\APACrefatitle {DECODE: a new method for discovering clusters of
  different densities in spatial data} {Decode: a new method for discovering
  clusters of different densities in spatial data}.{\BBCQ}
\newblock
\APACjournalVolNumPages{Data Mining and Knowledge Discovery}{18}{3}{337}.
\newblock
\begin{APACrefDOI} \doi{10.1007/s10618-008-0120-3} \end{APACrefDOI}
\PrintBackRefs{\CurrentBib}

\bibitem [\protect \citeauthoryear {%
Petersen%
\ \protect \BOthers {.}}{%
Petersen%
\ \protect \BOthers {.}}{%
{\protect \APACyear {2007}}%
}]{%
petersen2007}
\APACinsertmetastar {%
petersen2007}%
\begin{APACrefauthors}%
Petersen, I.%
, Johnson, A.%
, Islam, A.%
, Duckworth, G.%
, Livermore, D.%
\BCBL {}\ \BBA {} Hayward, A.%
\end{APACrefauthors}%
\unskip\
\newblock
\APACrefYearMonthDay{2007}{}{}.
\newblock
{\BBOQ}\APACrefatitle {Protective effect of antibiotics against serious
  complications of common respiratory tract infections: retrospective cohort
  study with the UK General Practice Research Database} {Protective effect of
  antibiotics against serious complications of common respiratory tract
  infections: retrospective cohort study with the uk general practice research
  database}.{\BBCQ}
\newblock
\APACjournalVolNumPages{Bmj}{335}{7627}{982}.
\newblock
\begin{APACrefDOI} \doi{10.1136/bmj.39345.405243.BE} \end{APACrefDOI}
\PrintBackRefs{\CurrentBib}

\bibitem [\protect \citeauthoryear {%
Povinelli%
, Johnson%
, Lindgren%
\BCBL {}\ \BBA {} Ye%
}{%
Povinelli%
\ \protect \BOthers {.}}{%
{\protect \APACyear {2004}}%
}]{%
povinelli2004}
\APACinsertmetastar {%
povinelli2004}%
\begin{APACrefauthors}%
Povinelli, R\BPBI J.%
, Johnson, M\BPBI T.%
, Lindgren, A\BPBI C.%
\BCBL {}\ \BBA {} Ye, J.%
\end{APACrefauthors}%
\unskip\
\newblock
\APACrefYearMonthDay{2004}{}{}.
\newblock
{\BBOQ}\APACrefatitle {Time series classification using Gaussian mixture models
  of reconstructed phase spaces} {Time series classification using gaussian
  mixture models of reconstructed phase spaces}.{\BBCQ}
\newblock
\APACjournalVolNumPages{IEEE Transactions on Knowledge and Data
  Engineering}{16}{6}{779--783}.
\newblock
\begin{APACrefDOI} \doi{10.1109/TKDE.2004.17} \end{APACrefDOI}
\PrintBackRefs{\CurrentBib}

\bibitem [\protect \citeauthoryear {%
Rodriguez%
\ \BBA {} Laio%
}{%
Rodriguez%
\ \BBA {} Laio%
}{%
{\protect \APACyear {2014}}%
}]{%
rodriguez2014}
\APACinsertmetastar {%
rodriguez2014}%
\begin{APACrefauthors}%
Rodriguez, A.%
\BCBT {}\ \BBA {} Laio, A.%
\end{APACrefauthors}%
\unskip\
\newblock
\APACrefYearMonthDay{2014}{}{}.
\newblock
{\BBOQ}\APACrefatitle {Clustering by fast search and find of density peaks}
  {Clustering by fast search and find of density peaks}.{\BBCQ}
\newblock
\APACjournalVolNumPages{Science}{344}{6191}{1492--1496}.
\newblock
\begin{APACrefDOI} \doi{10.1126/science.1242072} \end{APACrefDOI}
\PrintBackRefs{\CurrentBib}

\bibitem [\protect \citeauthoryear {%
Ruiz%
, Spiliopoulou%
\BCBL {}\ \BBA {} Menasalvas%
}{%
Ruiz%
\ \protect \BOthers {.}}{%
{\protect \APACyear {2007}}%
}]{%
ruiz2007}
\APACinsertmetastar {%
ruiz2007}%
\begin{APACrefauthors}%
Ruiz, C.%
, Spiliopoulou, M.%
\BCBL {}\ \BBA {} Menasalvas, E.%
\end{APACrefauthors}%
\unskip\
\newblock
\APACrefYearMonthDay{2007}{}{}.
\newblock
{\BBOQ}\APACrefatitle {C-dbscan: Density-based clustering with constraints}
  {C-dbscan: Density-based clustering with constraints}.{\BBCQ}
\newblock
\BIn{} \APACrefbtitle {International Workshop on Rough Sets, Fuzzy Sets, Data
  Mining, and Granular-Soft Computing} {International workshop on rough sets,
  fuzzy sets, data mining, and granular-soft computing}\ (\BPGS\ 216--223).
\newblock
\begin{APACrefDOI} \doi{10.1007/978-3-540-72530-5_25} \end{APACrefDOI}
\PrintBackRefs{\CurrentBib}

\bibitem [\protect \citeauthoryear {%
Schaeffer%
}{%
Schaeffer%
}{%
{\protect \APACyear {2007}}%
}]{%
schaeffer2007}
\APACinsertmetastar {%
schaeffer2007}%
\begin{APACrefauthors}%
Schaeffer, S\BPBI E.%
\end{APACrefauthors}%
\unskip\
\newblock
\APACrefYearMonthDay{2007}{}{}.
\newblock
{\BBOQ}\APACrefatitle {Graph clustering} {Graph clustering}.{\BBCQ}
\newblock
\APACjournalVolNumPages{Computer science review}{1}{1}{27--64}.
\newblock
\begin{APACrefDOI} \doi{10.1016/j.cosrev.2007.05.001} \end{APACrefDOI}
\PrintBackRefs{\CurrentBib}

\bibitem [\protect \citeauthoryear {%
Schroff%
, Kalenichenko%
\BCBL {}\ \BBA {} Philbin%
}{%
Schroff%
\ \protect \BOthers {.}}{%
{\protect \APACyear {2015}}%
}]{%
schroff2015}
\APACinsertmetastar {%
schroff2015}%
\begin{APACrefauthors}%
Schroff, F.%
, Kalenichenko, D.%
\BCBL {}\ \BBA {} Philbin, J.%
\end{APACrefauthors}%
\unskip\
\newblock
\APACrefYearMonthDay{2015}{}{}.
\newblock
{\BBOQ}\APACrefatitle {Facenet: A unified embedding for face recognition and
  clustering} {Facenet: A unified embedding for face recognition and
  clustering}.{\BBCQ}
\newblock
\BIn{} \APACrefbtitle {Proceedings of the IEEE conference on computer vision
  and pattern recognition} {Proceedings of the ieee conference on computer
  vision and pattern recognition}\ (\BPGS\ 815--823).
\PrintBackRefs{\CurrentBib}

\bibitem [\protect \citeauthoryear {%
Scott%
}{%
Scott%
}{%
{\protect \APACyear {1992}}%
}]{%
scott1992}
\APACinsertmetastar {%
scott1992}%
\begin{APACrefauthors}%
Scott, D\BPBI W.%
\end{APACrefauthors}%
\unskip\
\newblock
\APACrefYear{1992}.
\newblock
\APACrefbtitle {Multivariate Density Estimation. Theory, Practice and
  Visualization} {Multivariate density estimation. theory, practice and
  visualization}.
\newblock
\APACaddressPublisher{}{New York: Wiley}.
\PrintBackRefs{\CurrentBib}

\bibitem [\protect \citeauthoryear {%
Scott%
}{%
Scott%
}{%
{\protect \APACyear {2009}}%
}]{%
scott2009}
\APACinsertmetastar {%
scott2009}%
\begin{APACrefauthors}%
Scott, D\BPBI W.%
\end{APACrefauthors}%
\unskip\
\newblock
\APACrefYearMonthDay{2009}{}{}.
\newblock
{\BBOQ}\APACrefatitle {Sturges' rule} {Sturges' rule}.{\BBCQ}
\newblock
\APACjournalVolNumPages{Wiley Interdisciplinary Reviews: Computational
  Statistics}{1}{3}{303--306}.
\newblock
\begin{APACrefDOI} \doi{10.1002/wics.35} \end{APACrefDOI}
\PrintBackRefs{\CurrentBib}

\bibitem [\protect \citeauthoryear {%
Serra%
\ \BBA {} Arcos%
}{%
Serra%
\ \BBA {} Arcos%
}{%
{\protect \APACyear {2014}}%
}]{%
serra2014}
\APACinsertmetastar {%
serra2014}%
\begin{APACrefauthors}%
Serra, J.%
\BCBT {}\ \BBA {} Arcos, J\BPBI L.%
\end{APACrefauthors}%
\unskip\
\newblock
\APACrefYearMonthDay{2014}{}{}.
\newblock
{\BBOQ}\APACrefatitle {An empirical evaluation of similarity measures for time
  series classification} {An empirical evaluation of similarity measures for
  time series classification}.{\BBCQ}
\newblock
\APACjournalVolNumPages{Knowledge-Based Systems}{67}{}{305--314}.
\newblock
\begin{APACrefDOI} \doi{10.1016/j.knosys.2014.04.035} \end{APACrefDOI}
\PrintBackRefs{\CurrentBib}

\bibitem [\protect \citeauthoryear {%
Silverman%
}{%
Silverman%
}{%
{\protect \APACyear {1986}}%
}]{%
silverman1986}
\APACinsertmetastar {%
silverman1986}%
\begin{APACrefauthors}%
Silverman, B.%
\end{APACrefauthors}%
\unskip\
\newblock
\APACrefYearMonthDay{1986}{}{}.
\newblock
{\BBOQ}\APACrefatitle {Monographs on statistics and applied probability}
  {Monographs on statistics and applied probability}.{\BBCQ}
\newblock
\APACjournalVolNumPages{Density estimation for statistics and data
  analysis}{26}{}{}.
\PrintBackRefs{\CurrentBib}

\bibitem [\protect \citeauthoryear {%
Smith%
, Lessells%
, Grant%
, Herbst%
\BCBL {}\ \BBA {} Tanser%
}{%
Smith%
\ \protect \BOthers {.}}{%
{\protect \APACyear {2018}}%
}]{%
smith2018}
\APACinsertmetastar {%
smith2018}%
\begin{APACrefauthors}%
Smith, C.%
, Lessells, R.%
, Grant, A.%
, Herbst, K.%
\BCBL {}\ \BBA {} Tanser, F.%
\end{APACrefauthors}%
\unskip\
\newblock
\APACrefYearMonthDay{2018}{}{}.
\newblock
{\BBOQ}\APACrefatitle {Spatial clustering of drug-resistant tuberculosis in
  Hlabisa subdistrict, KwaZulu-Natal, 2011--2015} {Spatial clustering of
  drug-resistant tuberculosis in hlabisa subdistrict, kwazulu-natal,
  2011--2015}.{\BBCQ}
\newblock
\APACjournalVolNumPages{The international journal of tuberculosis and lung
  disease}{22}{3}{287--293}.
\newblock
\begin{APACrefDOI} \doi{10.5588/ijtld.17.0457} \end{APACrefDOI}
\PrintBackRefs{\CurrentBib}

\bibitem [\protect \citeauthoryear {%
Sugiyama%
}{%
Sugiyama%
}{%
{\protect \APACyear {2015}}%
}]{%
sugiyama2015}
\APACinsertmetastar {%
sugiyama2015}%
\begin{APACrefauthors}%
Sugiyama, M.%
\end{APACrefauthors}%
\unskip\
\newblock
\APACrefYear{2015}.
\newblock
\APACrefbtitle {Introduction to statistical machine learning} {Introduction to
  statistical machine learning}.
\newblock
\APACaddressPublisher{}{Morgan Kaufmann}.
\PrintBackRefs{\CurrentBib}

\bibitem [\protect \citeauthoryear {%
Terrell%
, Scott%
\BCBL {}\ \protect \BOthers {.}}{%
Terrell%
\ \protect \BOthers {.}}{%
{\protect \APACyear {1992}}%
}]{%
terrell1992}
\APACinsertmetastar {%
terrell1992}%
\begin{APACrefauthors}%
Terrell, G\BPBI R.%
, Scott, D\BPBI W.%
\BCBL {}\ \BOthersPeriod {.}\end{APACrefauthors}%
\unskip\
\newblock
\APACrefYearMonthDay{1992}{}{}.
\newblock
{\BBOQ}\APACrefatitle {Variable kernel density estimation} {Variable kernel
  density estimation}.{\BBCQ}
\newblock
\APACjournalVolNumPages{The Annals of Statistics}{20}{3}{1236--1265}.
\newblock
\begin{APACrefDOI} \doi{10.1214/aos/1176348768} \end{APACrefDOI}
\PrintBackRefs{\CurrentBib}

\bibitem [\protect \citeauthoryear {%
Tormene%
, Giorgino%
, Quaglini%
\BCBL {}\ \BBA {} Stefanelli%
}{%
Tormene%
\ \protect \BOthers {.}}{%
{\protect \APACyear {2009}}%
}]{%
tormene2009}
\APACinsertmetastar {%
tormene2009}%
\begin{APACrefauthors}%
Tormene, P.%
, Giorgino, T.%
, Quaglini, S.%
\BCBL {}\ \BBA {} Stefanelli, M.%
\end{APACrefauthors}%
\unskip\
\newblock
\APACrefYearMonthDay{2009}{}{}.
\newblock
{\BBOQ}\APACrefatitle {Matching incomplete time series with dynamic time
  warping: an algorithm and an application to post-stroke rehabilitation}
  {Matching incomplete time series with dynamic time warping: an algorithm and
  an application to post-stroke rehabilitation}.{\BBCQ}
\newblock
\APACjournalVolNumPages{Artificial intelligence in medicine}{45}{1}{11--34}.
\newblock
\begin{APACrefDOI} \doi{10.1016/j.artmed.2008.11.007} \end{APACrefDOI}
\PrintBackRefs{\CurrentBib}

\bibitem [\protect \citeauthoryear {%
Williams%
}{%
Williams%
}{%
{\protect \APACyear {2011}}%
}]{%
williams2011}
\APACinsertmetastar {%
williams2011}%
\begin{APACrefauthors}%
Williams, E.%
\end{APACrefauthors}%
\unskip\
\newblock
\APACrefYearMonthDay{2011}{}{}.
\newblock
{\BBOQ}\APACrefatitle {Aviation Formulary V1. 42} {Aviation formulary v1.
  42}.{\BBCQ}
\newblock
\APACjournalVolNumPages{Aviation}{1}{}{42}.
\newblock
\begin{APACrefURL}
  \url{ftp://ftp.bartol.udel.edu/anita/amir/My_thesis/Figures4Thesis/CRC_plots/Aviation%20Formulary%20V1.46.pdf}
  \end{APACrefURL}
\PrintBackRefs{\CurrentBib}

\bibitem [\protect \citeauthoryear {%
Xing%
, Pei%
\BCBL {}\ \BBA {} Philip%
}{%
Xing%
\ \protect \BOthers {.}}{%
{\protect \APACyear {2012}}%
}]{%
xing2012}
\APACinsertmetastar {%
xing2012}%
\begin{APACrefauthors}%
Xing, Z.%
, Pei, J.%
\BCBL {}\ \BBA {} Philip, S\BPBI Y.%
\end{APACrefauthors}%
\unskip\
\newblock
\APACrefYearMonthDay{2012}{}{}.
\newblock
{\BBOQ}\APACrefatitle {Early classification on time series} {Early
  classification on time series}.{\BBCQ}
\newblock
\APACjournalVolNumPages{Knowledge and information systems}{31}{1}{105--127}.
\newblock
\begin{APACrefDOI} \doi{10.1007/s10115-011-0400-x} \end{APACrefDOI}
\PrintBackRefs{\CurrentBib}

\bibitem [\protect \citeauthoryear {%
Zelnik-Manor%
\ \BBA {} Perona%
}{%
Zelnik-Manor%
\ \BBA {} Perona%
}{%
{\protect \APACyear {2005}}%
}]{%
zelnik2005}
\APACinsertmetastar {%
zelnik2005}%
\begin{APACrefauthors}%
Zelnik-Manor, L.%
\BCBT {}\ \BBA {} Perona, P.%
\end{APACrefauthors}%
\unskip\
\newblock
\APACrefYearMonthDay{2005}{}{}.
\newblock
{\BBOQ}\APACrefatitle {Self-tuning spectral clustering} {Self-tuning spectral
  clustering}.{\BBCQ}
\newblock
\BIn{} \APACrefbtitle {Advances in neural information processing systems}
  {Advances in neural information processing systems}\ (\BPGS\ 1601--1608).
\PrintBackRefs{\CurrentBib}

\end{thebibliography}
\bibliographystyle{apacite}

\label{References}

\end{document}